\documentclass[openacc]{rstransa}

\jname{rsta}
\Journal{Phil. Trans. R. Soc.}

\newcommand{\h}{\hspace{6mm}}

\newcommand{\beq}{\begin{equation}}
\newcommand{\eeq}{\end{equation}}
\newcommand{\beqa}{\begin{eqnarray}}
\newcommand{\eeqa}{\end{eqnarray}}

\begin{document}

\title{Fast ion shuttling which is  robust versus oscillatory perturbations}

\author{
H. Espin\'os$^{1}$, J. Echanobe$^{2}$, X.-J- Lu$^{3}$ and J. G. Muga$^{1}$}

\address{$^{1}$Department of Physical Chemistry, University of the Basque Country UPV/EHU, Apdo 644, Bilbao, Spain\\
$^{2}$Departamento de Electricidad y Electrónica, UPV/EHU, Apdo 644, Bilbao, Spain\\
$^{3}$School of Electric and Mechatronics Engineering, Xuchang University, Xuchang 461000, China}

\subject{quantum physics, quantum engineering}

\keywords{particle shuttling, shortcuts to adiabaticity}

\corres{J. G. Muga\\
\email{jg.muga@ehu.es}}

\begin{abstract}
Shuttling protocols designed by shortcut-to-adiabaticity techniques may suffer from perturbations and imperfect implementations. 
We study the motional excitation of a single ion shuttled in harmonic traps with time-dependent, ``systematic''
oscillatory perturbations around the nominal parameters. These elementary perturbations could form any other by superposition. 
Robust shuttling strategies are proposed and compared, and optimizations are performed. 
\end{abstract}


\begin{fmtext}

\section{Introduction}
Shuttling one or several atoms or ions is a key operation for fundamental research and to implement quantum-based technologies. 
In many of these applications it is important to deliver the particles fast and motionally unexcited at destination. 
Shortcut-to-adiabaticity (STA) techniques  \cite{Torrontegui2013,Guery2019} provide transport protocols for that end, but in practice the nominal trajectory of the control parameters 
is not implemented exactly because of technical  imperfections and constraints. These control errors  pose the need for a) studying the effect
of perturbations on STA-based protocols and b) devising protocols that are robust with respect to imperfections and satisfy the technical constraints. 

Noisy perturbations in STA-based shuttling operations have been studied quite thoroughly recently  \cite{Lu2020}. 
In this work we shall address a complementary aspect, namely the effect of ``systematic''  oscillatory perturbations of the ideal trap frequency or of the trap trajectory, and design robust protocols. 
\end{fmtext}
\maketitle
The  intermediate regime of excitations between an elementary  monochromatic perturbation and 
a noisy one could be handled by linearly combining monochromatic perturbations, but a full understanding of the effect of a monochromatic 
perturbation is needed first.   
The theoretical analysis is done for one single particle and it is quite general within the harmonic oscillator assumption for the trap, 
but numerical examples and physical motivation for approximations and parameter values are borrowed from realistic trapped ion settings.

There are several STA approaches to shuttle  a single particle or condensate from 
the trap position $Q(0)=0$ at $t=0$ to $Q(T)=d$ in a transport time $T$  \cite{Muga2021}. 
Invariant-based inverse engineering and the ``Fourier method''  have been widely used the design the trap motion 
and will be the  core approaches here.

 In particular, the transport in a harmonic trap may be inverse engineered using quadratic invariants of motion \cite{Torrontegui2011} (alternatively ``scaling'' for condensates \cite{Muga2009,Schaff2011,Torrontegui2012}).  The invariant eigenvectors are also very useful to describe the dynamics, and combined with perturbation theory, they provide compact expressions for the energy excitation.

If shuttling is performed in a rigid harmonic oscillator,  with constant trap frequency, and vanishing trap speed  at initial and final times, the final excitation energy can also be expressed in terms of  the Fourier transform (FT) of the trap acceleration (or velocity)
at the trap frequency \cite{Landau1976,Bowler2012,Couvert2008,Guery2014,Reichle2006,MartinezCercos2020}. 
A consequence is that if the ideal, excitation-free trap trajectory is affected by some perturbation or deviation, the final excitation only depends on the Fourier transform of the {\it deviation} of the trap acceleration with respect to the ideally designed one. These deviations may be independent 
of the ideal trajectory, for example if they are due to homogeneous background noise, or may depend on it, e.g. if some locations 
are more prone to errors.  This possible dependence makes in general,  smooth, band-limited 
and spatially-limited ideal trajectories
preferable.   
A formal framework to design trap trajectories to nullify the FT of the acceleration at the trap frequency  can  be worked out systematically \cite{Guery2014,MartinezCercos2020}, without making explicit use of invariants. 
In fact combining invariants and the Fourier forms as we do in this work is worthwhile.  
In particular,  whereas the Fourier method, as used so far, needs constant trap frequencies,  here  we shall extend it to time-dependent 
perturbations and apply optimization strategies.

%
%
%
%

In Section \ref{invsec}, we start by applying invariant-based inverse engineering to shuttling, 
modifying the perturbative treatment developed for noisy perturbations, to determine the effect of arbitrary perturbations. We also find compact FT expressions of the 
excitation; 
In Section \ref{polsec}, we apply the previous general results to harmonic transport with an elementary  sinusoidal perturbation in the trap frequency. We focus on a specific polynomial STA protocol and study the different contributions to the final energy;
In Section \ref{optsec}, we employ several techniques to find trap trajectories that satisfy different optimization criteria when the trap frequency is affected by one or more sinusoidal perturbations;
The conclusions are presented in Section \ref{consec}.
Throughout the work we shall assume an effective one-dimensional transport, which is realistic with current experimental settings.
\section{Transport of an ion using invariant-based inverse engineering\label{invsec}}
The Hamiltonian of a particle of mass $m$  in a harmonic oscillator of (angular) frequency $\Omega(t)$,  shuttled along  $Q(t)$,
\begin{equation}\label{eq:hamiltonian}
H(t)=\frac{p^2}{2m}+\frac{m\Omega^2(t)}{2}\big[x-Q(t)\big]^2,
\end{equation}
has a quadratic Lewis-Riesenfeld invariant \cite{Lu2020}
\beqa
I(t)&=&\frac{1}{2m}\bigg\{\rho(t)\big[p-m\dot{q}_c(t)\big]-m\dot{\rho}(t)\big[x-q_c(t)\big]\bigg\}^2
+\frac{1}{2}m\Omega_0^2\bigg[\frac{x-q_c(t)}{\rho(t)}\bigg]^2,
\eeqa
where $\rho(t)$ is a scaling factor for the width of the eigenstates of $I$, and $q_c(t)$ is a classical trajectory for the forced oscillator. The dots denote time derivatives. The invariant satisfies
\begin{equation}
\label{invar}
\frac{dI(t)}{dt}=\frac{\partial{I(t)}}{\partial{t}}+\frac{1}{i\hbar}\big[I(t),H(t)\big]=0,
\end{equation}
so that its expectation value for states driven by \textit{H(t)} is constant. From  (\ref{invar}), we find the "Ermakov" and "Newton" equations,
\begin{eqnarray}\label{eq:Ermakov}
\ddot{\rho}(t)+\Omega^2(t)\rho(t)&=&\frac{\Omega_0^2}{\rho^3(t)},
\\\label{eq:Newton}
\ddot{q}_c(t)+\Omega^2(t)q_c(t)&=&\Omega^2(t)Q(t).
\end{eqnarray}
Hereafter we choose for convenience $\Omega_0=\Omega(0)$. The main idea for inverse engineering a quiet driving 
is to design $q_c(t)$ and introduce it in  (\ref{eq:Newton}) to deduce special trap trajectories  without final excitation. We impose the initial conditions
\begin{eqnarray}
q_c(0)=0,  &&\hspace{1.32cm}\rho(0)=1, \nonumber\\\label{eq:6}
\dot{q}_c(0)=0,&&\hspace{1.32cm}\dot{\rho}(0)=0,\\
\ddot{q}_c(0)=0,&&\hspace{1.32cm}\ddot{\rho}(0)=0, \nonumber
\end{eqnarray} 
so that the invariant commutes with the Hamiltonian at $t=0$. The last two guarantee the continuity of $Q(t)$ and $\Omega(t)$ at the initial time. Similar conditions (except for $q_c(T)=d$) are needed at final time $T$ to achieve  excitationless shuttling. We shall use a specific notation, $Q_0(t)$, for the ideal trap trajectory deduced from Newton's equation for the $q_c(t)$ that satisfy the imposed boundary conditions when $\Omega(t)=\Omega_0$. 
The actual, experimentally implemented trap trajectory, $Q(t)$, and the actual trap frequency may differ from $Q_0(t)$
and $\Omega_0$  producing  motional excitation at time $T$. 
\subsection{Final and transient energies}
Let us calculate the final energy due to deviations in the ideal trap trajectory and trap frequency. Now we assume that $Q(t)$ and $\Omega(t)$
are given, and $q_c(t)$ and $\rho(t)$ are found from them, using (\ref{eq:Ermakov}) and (\ref{eq:Newton}) and  the initial conditions (\ref{eq:6}). 
The wave-function is found through the Lewis-Riesenfeld invariant. The corresponding eigenstates can be found analytically \cite{Lu2020},
\begin{equation}
\psi_n(x,t)=\frac{1}{\sqrt{\rho}}e^{\frac{im}{\hbar}\left[\frac{\dot{\rho}x^2}{2\rho}+\frac{\left(\dot{q}_c\rho-\dot{\rho}q_c\right)x}{\rho}\right]}\phi_n\left(\frac{x-q_c}{\rho}\right),
\end{equation}
where $\phi_n$ is the $n$-th eigenstate of the rigid harmonic oscillator of frequency $\Omega_0$. Elementary solutions of the time-dependent Schrödinger equation may be written as 
$$
\Psi_n(x,t)=e^{i\theta_n(t)}\psi_n(x,t),
$$
where $\theta_n(t)$ are the Lewis-Riesenfeld phases, which are found so that $\Psi_n$ is indeed a solution.
%
%
At final time $T$, the  harmonic trap is in $Q(T)$ and its frequency is $\Omega(T)$, not necessarily equal to $d$ and $\Omega_0$, respectively- 
The final energy can be found exactly as
\begin{eqnarray}
E_n(T)&=&\big\langle H(T)\big\rangle=\bigg\langle\frac{p^2}{2m}+\frac{m\Omega^2(T)}{2}(x-Q(T))^2\bigg\rangle
\nonumber\\
\label{eq:energy}
&=&\frac{m\Omega^2(T)}{2}\left[q_c(T)-Q(T)\right]^2+\frac{m}{2}\dot{q}_c^2(T)\nonumber\\
&+&\frac{\hbar}{4\Omega_0}(2n+1)\left[\dot{\rho}^2(T)+\frac{\Omega_0^2}{\rho^2(T)}+\Omega^2(T)\rho^2(T)\right]\!.
\end{eqnarray}
Some terms depend on the trap trajectory $Q(t)$ (also through $q_c(t)$), while others do not. Following Lu {\it et al.}  \cite{Lu2020}, {\em we call the trap-motion independent terms ``static", and the dependent ones, ``dynamical"}.

Equation (\ref{eq:energy}) is also  valid with the change $T\rightarrow t$ for any time $t$ during the transport. 
%
%
%
%
%
%
\subsection{Perturbation in the trap frequency\label{section:2.2}}
Assume that the trap frequency is perturbed as 
\begin{equation}\label{eq:12}
\Omega(t)=\Omega_0\left[1+\lambda f(t)\right],
\end{equation}
where $\lambda$ is  a dimensionless perturbative parameter much smaller than 1 during the calculations, and $f(t)$ can be any (dimensionless) function. 
We assume now  $Q(t)=Q_0(t)$, with $Q_0(0)=0$ and $Q_0(T)=d$.

To analyze the effect of the  perturbation, $\rho(t)$ and $q_c(t)$ are expanded in powers of $\lambda$,
\begin{eqnarray}\label{eq:13}
\rho(t)&=&\rho^{(0)}(t)+\lambda\rho^{(1)}(t)+O(\lambda^2),
\nonumber\\
q_c(t)&=&q_c^{(0)}(t)+\lambda q_c^{(1)}(t)+O(\lambda^2).
\end{eqnarray}
The zeroth order, or unperturbed limit, corresponding to $\Omega(t)=\Omega_0$, fulfills
\begin{eqnarray}\label{eq:14}
\rho^{(0)}(t)=1,\;\;\;\;
\ddot{q}_c^{(0)}(t)+\Omega_0^2q_c^{(0)}(t)=\Omega_0^2Q_0(t),
\end{eqnarray}
with $q_c^{(0)}$ satisfying the initial conditions (\ref{eq:6}). To minimize the final excitation and make $Q_0(t)$ continuous at $t=T$, also the following boundary conditions have to be imposed,
\begin{eqnarray}\label{eq:15}
q^{(0)}_c(T)=d,\;\;\;\; 
\dot{q}^{(0)}_c(T)=
\ddot{q}^{(0)}_c(T)=0.
\end{eqnarray}
With these conditions, the final energy given by  (\ref{eq:energy}) is simply the $n$-th energy level of the static harmonic oscillator with the unperturbed frequency $\Omega_0$,
\begin{equation}
E_n^{(0)}=\hbar\Omega_0(n+1/2).
\end{equation}
If  $q_c^{(0)}(t)$ is designed with the aforementioned boundary conditions,  $Q_0(t)$ is found from  (\ref{eq:14}), and excitationless transport is guaranteed  in the unperturbed limit. Introducing the  expansions for $\Omega(t)$, $\rho(t)$ and $q_c(t)$ given by  (\ref{eq:12})  and (\ref{eq:13}), into  (\ref{eq:energy}), we find an expansion of the final energy with dynamical and static contributions.
%
Combining the zeroth and first order in $\lambda$ of both the dynamical and the static terms, we get 
\beq
E_n^{(0)}+\lambda E_n^{(1)}=\hbar\Omega(T)\left(n+1/2\right).
\eeq
For the second order in  $E_n=E_n^{(0)}+\lambda E_n^{(1)}+\lambda^2 E_n^{(2)}+...$, 
the dynamical and static terms are
\begin{eqnarray}\label{eq:energy2}
E_n^{(2)}(T)&=&\frac{m\Omega_0^2}{2}\left\{\left[q_c^{(1)}(T)\right]^2+\frac{1}{\Omega_0^2}\left[\dot{q}_c^{(1)}(T)\right]^2\right\} \nonumber\\
&+& \frac{\hbar\Omega_0}{4}(2n+1)\!\left\{\!\left[2\rho^{(1)}(T)+f(T)\right]^2\!+\!\frac{\left[\dot{\rho}^{(1)}(T)\right]^2}{\Omega_0^2}\!\right\}\!.
\end{eqnarray}
To achieve robust shuttling protocols, the main goal is to minimize this last expression. Substituting the expansions of $\rho(t)$ and $q_c(t)$ into  (\ref{eq:Ermakov}) and (\ref{eq:Newton}), we  find the differential equations satisfied by $\rho^{(1)}(t)$ and $q_c^{(1)}(t)$ by keeping only the first order in $\lambda$,
%
	%
	\begin{equation}\label{eq:17}
	\ddot{\rho}^{(1)}(t)+4\Omega_0^2\rho^{(1)}(t)=-2\Omega_0^2f(t),
	\end{equation}
	with initial conditions $\rho^{(1)}(0)=\dot{\rho}^{(1)}(0)=0$, and 
%
	%
	\begin{equation}\label{eq:18}
	\ddot{q}_c^{(1)}(t)+\Omega_0^2q_c^{(1)}(t)=2f(t)\ddot{q}_c^{(0)}(t),
	\end{equation}
	with initial conditions $q_c^{(1)}(0)=\dot{q}_c^{(1)}(0)=0$.
%
Equations (\ref{eq:17}) and (\ref{eq:18}) admit a formal solution,
\begin{eqnarray}\label{eq:19}
\rho^{(1)}(t)&=&-\Omega_0\int_0^tdt'f(t')\sin\left[2\Omega_0(t-t')\right],
\\ \label{eq:20}
q_c^{(1)}(t)&=&\frac{2}{\Omega_0}\int_0^tdt'f(t')\ddot{q}_c^{(0)}(t')\sin\left[\Omega_0(t-t')\right],
\end{eqnarray}
with similar expressions for their first time derivatives.
\subsection{Perturbation in the trap trajectory}\label{sect:2.3}
Complementing the previous section, consider now a constant trap frequency $\Omega_0$, but errors in the trap position, 
\beq\label{eq:q0}
Q(t)=Q_0(t)+\varepsilon dh(t),
\eeq
where $\varepsilon$ is  the dimensionless perturbative parameter, and $h(t)$ is a (dimensionless) time dependent function. The distance $d$ is included to the set the scale and make the expression dimensionally consistent. For neutral atom transport in optical lattices iterative approaches to minimize deviations have been put forward \cite{Lam2021}. 
In the numerical voltage optimization performed in the trapped ion laboratories, there is some choice on whether minimizing deviations of the trap frequency or the trap position, see e.g \cite{Muga2021}. We shall discuss later where the emphasis has to be put on. Indeed, in a trapped ion experiment, perturbations 
cannot be suppressed to any desired level because of technical imperfections and limitations of the  control, for example the voltages have upper limits, the time resolution is limited, the number of electrodes is limited, and their geometry is fixed \cite{Muga2021}.  In this context advice on where to put the emphasis in parameter optimizations is quite useful.

As in  (\ref{eq:13}), we expand  $\rho(t)$ and $q_c(t)$ with 
the change $\lambda\to\epsilon$.
We introduce these expansions, together with (\ref{eq:q0}) for $Q(t)$, into Ermakov and Newton equations. The zeroth order energy fulfills once again (\ref{eq:14}),
with $q_c^{(0)}$ satisfying the boundary conditions (\ref{eq:6}) and (\ref{eq:15}). The first order is zero and 
for the second order of the final excitation we get 
\begin{eqnarray}\label{eq:energy3}
E_n^{(2)}(T)&=&\frac{m\Omega_0^2}{2}\left\{\left[q_c^{(1)}(T)-dh(T)\right]^2+\frac{1}{\Omega_0^2}\left[\dot{q}_c^{(1)}(T)\right]^2\right\} \nonumber\\
&+& \frac{\hbar\Omega_0}{4}(2n+1)\left\{4\left[\rho^{(1)}(T)\right]^2+\frac{1}{\Omega_0^2}\left[\dot{\rho}^{(1)}(T)\right]^2\right\}.
\end{eqnarray}
$\rho^{(1)}(t)$ and $q_c^{(1)}(t)$ satisfy 
\begin{eqnarray}
\ddot{\rho}^{(1)}(t)+4\Omega_0^2\rho^{(1)}(t)&=&0,\nonumber \\
\ddot{q}_c^{(1)}(t)+\Omega_0^2q_c^{(1)}(t)&=&d\Omega_0^2h(t),
\end{eqnarray}
with initial conditions $\rho^{(1)}(0)=\dot{\rho}^{(1)}(0)=0$ and $q_c^{(1)}(0)=\dot{q}_c^{(1)}(0)=0$. The solutions are
\begin{eqnarray}
\rho^{(1)}(t)&=&0,\label{eq:2.30} \\ 
q_c^{(1)}(t)&=&d\Omega_0\int_0^tdt'h(t')\sin\left[\Omega_0(t-t')\right].\label{eq:2.31}
\end{eqnarray}
Neither of the auxiliary variables depends on the trap trajectory, $Q_0(t)$, so there is only a static contribution to the second-order 
excitation\footnote{We assume that $h(t)$ does not depend on $q_c^{(0)}(t)$ or $Q_0(t)$}. For fixed $T$ it is not possible to design an optimal trap trajectory that minimizes the excitation. To diminish the effect of a perturbation in the trap position we may choose $T$ to make  (\ref{eq:2.31}) and its derivative zero at $T$. This may be done systematically if the form of the perturbation function $h(t)$ is known as we shall see.
\subsection{The Fourier forms}\label{section:2.4}
Here we find compact expressions 
for the excitation energy in the form of Fourier transforms. 
\subsubsection{Perturbation in the trap frequency}\label{section:2.4a}
Let us start by rewriting the term that depends on the classical trajectory $q_c(t)$ and its time derivative ---the dynamical term in (\ref{eq:energy2})-- 
as 
\begin{eqnarray}
E_{n,dynamical}^{(2)}=
\frac{m\Omega_0^2}{2}\Big\arrowvert q_c^{(1)}(T)-\frac{i}{\Omega_0}\dot{q}_c^{(1)}(T)\Big\arrowvert^2.
\end{eqnarray}
Now, we introduce the integral expressions for $q_c^{(1)}(t)$ and $\dot{q}_c^{(1)}(t)$, see   (\ref{eq:20}),
to write 
\begin{equation}\label{eq:2.39}
E_{n,dynamical}^{(2)}=2m\left\arrowvert\int_0^Tdt\,f(t)\ddot{q}_c^{(0)}(t)e^{-i\Omega_0t}\right\arrowvert^2.
\end{equation}
The same procedure can be applied to the static term (\ref{eq:energy2}). Assuming that there is no perturbation at final time, i.e., $\Omega(T)=\Omega_0$, or equivalently $f(T)=0$, we can write it  as 
\begin{eqnarray}
E^{(2)}_{n,static}=
\frac{\hbar\Omega_0}{4}(2n+1)\Big\arrowvert 2\rho^{(1)}(T)-\frac{i}{\Omega_0}\dot{\rho}^{(1)}(T)\Big\arrowvert^2.
\end{eqnarray}
Introducing now the integral expressions for $\rho^{(1)}(t)$ and $\dot{\rho}^{(1)}(t)$, see  (\ref{eq:19}), we get
\beq
E^{(2)}_{n,static}=\hbar\Omega_0^3(2n+1)\left\arrowvert\int_0^Tdt\,f(t)e^{-2i\Omega_0t}\right\arrowvert^2.\label{eq:2.40}
\eeq
\subsubsection{Perturbation in the trap position}
As discussed in  subsection \ref{invsec}\ref{sect:2.3}, the second order excitation due to a perturbation in the trap trajectory is purely static. When we apply the same kind of manipulations as before to this component, using  now  (\ref{eq:2.30}) into the final excitation (\ref{eq:energy3}), we get
\begin{equation}\label{eq:2.43}
E_{n,static}^{(2)}=\frac{m\Omega_0^4d^2}{2}\left\arrowvert\int_0^Tdt\,h(t)e^{-i\Omega_0t}\right\arrowvert^2,
\end{equation}
where we have assumed  $Q(T)=d$ and $h(T)=0$.

The static excitation  (\ref{eq:2.43}) is very similar to the one produced by a time-dependent deviation in the trap frequency, (\ref{eq:2.40}). 
Assuming that the perturbation functions $f(t)$ and $h(t)$ are similar, and that the parameters $\lambda$ and $\varepsilon$ are of the same order, there are mainly two differences between these two expressions. Firstly, the Fourier transform is evaluated at $2\Omega_0$  in  (\ref{eq:2.40}) 
and at $\Omega_0$  
in  (\ref{eq:2.43}).  Secondly, the prefactors  are different. Their ratio is
\begin{equation}
\eta_n=\frac{\hbar\Omega_0^3(2n+1)}{{m\Omega_0^4d^2}/{2}}=\frac{2\hbar(2n+1)}{m\Omega_0 d^2}.
\end{equation}
For the typical experimental values to shuttle an ion, 
this parameter is much smaller than 1. This means that in principle (for similar contributions of the moduli) {\em it is preferable to have an absolute 
control of the trap position  even if that compromises the control over the trap frequency}.
\section{Polynomial STA protocol for a transport with an oscillating trap frequency\label{polsec}}
We discussed in the previous section that special attention should be paid to perfectly adjusting the harmonic potential trajectory to the theoretically designed one, even if this implies assuming some errors in the trap frequency. In this section, we shall focus on trap frequency errors. While the static component is  the same for every STA trap trajectory with a given duration $T$, the dynamical one can be optimized with an appropriate trajectory. From now on, we consider a sinusoidal perturbation for the trap frequency,
\begin{equation}\label{eq:3.1}
\Omega(t)=\Omega_0\left[1+\lambda\sin(\omega t)\right],
\end{equation} 
\begin{figure}[h]
	\begin{center}
	\includegraphics[width=0.80\textwidth]{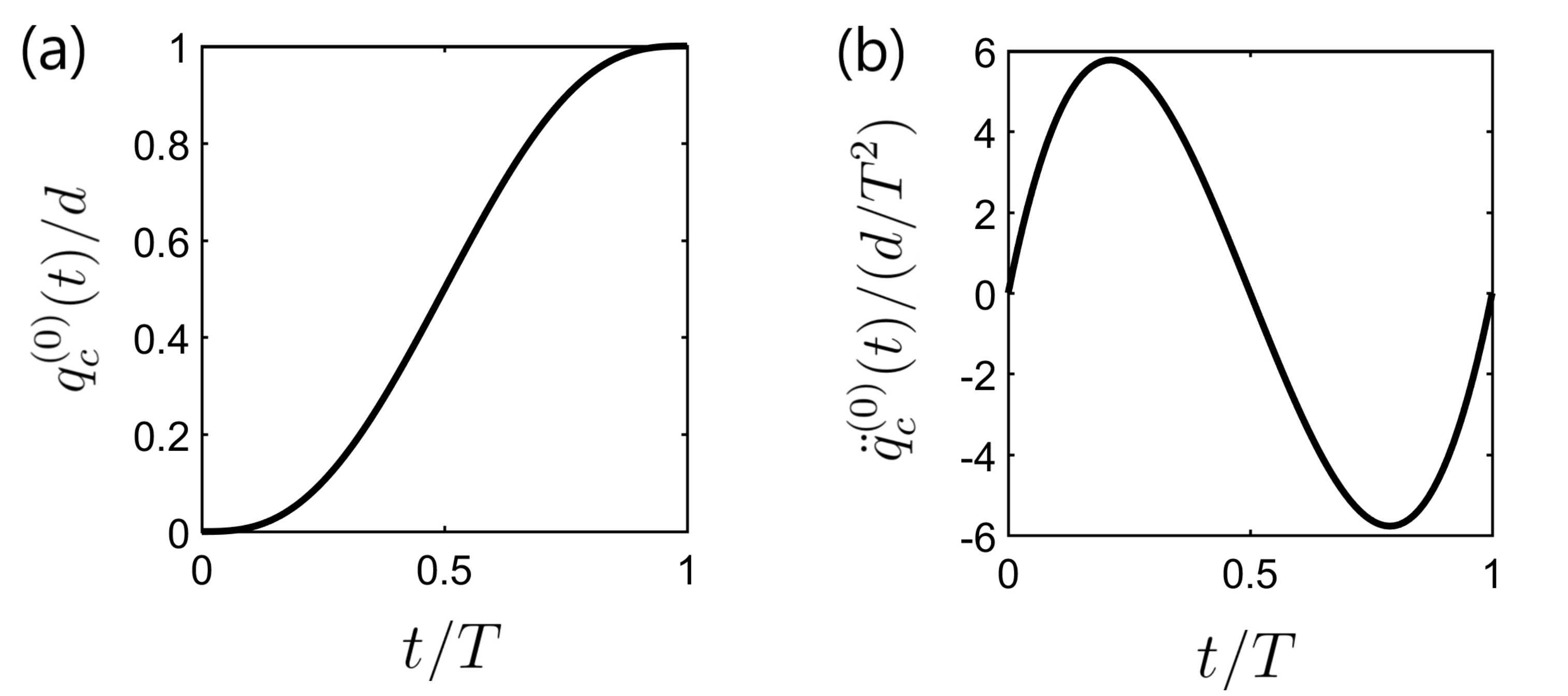}
		\caption{(a) Classical trajectory $q_c^{(0)}(t)/d$ versus $t/T$ and (b) acceleration of the classical trajectory $\ddot{q}_c(t)/(d/T^2)$ versus $t/T$ for the polynomial protocol.}
		\label{fig:3.1}
		\centering
	\end{center}
\end{figure}
to later consider the combination of several sines.  
This perturbation can be understood as an elementary Fourier component of an arbitrary perturbation. We can apply the results from the previous section, particularly from subsection \ref{invsec}\ref{section:2.2}, to this elementary perturbation.
We start by designing a classical trajectory $q_c^{(0)}(t)$ that satisfies the boundary conditions (\ref{eq:6}) and (\ref{eq:15}) as a 5th order polynomial,
\begin{equation}\label{eq:24}
q_c^{(0)}(t)=10d\left(\frac{t}{T}\right)^3-15d\left(\frac{t}{T}\right)^4+6d\left(\frac{t}{T}\right)^5,
\end{equation}
namely, the simplest polynomial that satisfies all six boundary conditions, and for that reason it has been used often \cite{Lu2020,Torrontegui2011,Zhang2016}. Once $q_c^{(0)}(t)$ is set (and so is the acceleration, see figure \ref{fig:3.1}), we get the trap trajectory $Q_0(t)$ from  (\ref{eq:14}). For short transport times $T$, $Q_0(t)$ could exceed the domain $[0,d]$. This  occurs symmetrically at both edges for $\Omega_0T\le2.505$ \cite{Torrontegui2011} (this value is independent of the total distance $d$). The excess beyond $[0,d]$  may be a problem in practice and a remedy will be discussed later on. 
\begin{figure}[ht]
	\begin{center}
\includegraphics[width=0.80\textwidth]{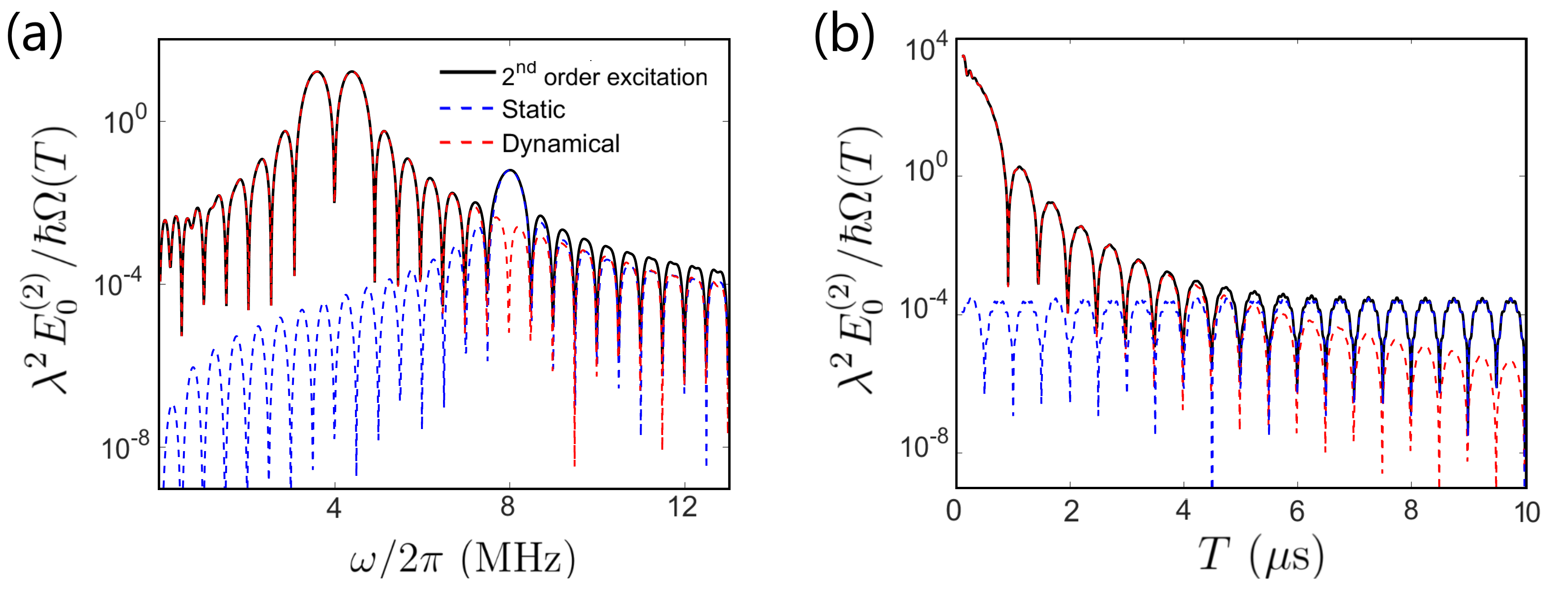}	
		\caption{(a) Log plot of the second order ground-state excitation due to a sinusoidal perturbation in the trap frequency (in units of final quanta) versus $\omega$. The transport is driven by a 5th order polynomial STA protocol. The black solid line is the total excitation, the blue dashed line is the static component and the red dashed line the dynamical component. The parameters are $\lambda=0.01$, $\Omega_0=2\pi\times4$ MHz, $m=1.455\cdot10^{-25}$ kg ($^{{88}}$Sr$^+$ ion), $d=50\;\mu$m and $T=2\;\mu$s.
(b) Log plot of the second order ground state excitation due to a sinusoidal perturbation in the trap frequency (in units of final quanta) versus $T$. Same parameters $\lambda$, $\Omega_0$, $m$, $d$, and color code as in (a), $\omega=2\pi\times 6$ MHz.	}
		\label{fig:1}
		\centering		
	\end{center}
\end{figure}
\subsection{Final energy using the perturbation method}
In figure \ref{fig:1}(a), the second order final excitation of a particle which is initially in its ground state, $E_0^{(2)}(T)$, and its two components, $static$ and $dynamical$, are shown versus  $\omega$, for a $^{88}$Sr$^+$ ion  shuttled a distance $d=50$ $\mu$m in $T=2$ $\mu$s using a trap with frequency $\Omega_0=2\pi\times 4$ MHz and the polynomial  (\ref{eq:24}), see details in caption (these values are realistic for current shuttling experiments). 

The dynamical component experiences a resonance at $\omega=\Omega_0$, and the static one at $\omega=2\Omega_0.$\footnote{We define these resonances phenomenolgically here, as the frequencies {\it around which} maximum excitation is found. They are better identified by the maximal envelope of the excitation rather than by the excitation itself. Note that the resonance at $2\Omega_0$ is a ``parameric resonance''.} In an experimental setting in which the trap frequency $\Omega_0$ is tunable and the perturbation frequency $\omega$ --or, at least, a dominant Fourier component of the perturbation-- is known, these resonances should be avoided. 

We also represent the two contributions to the excitation energy versus the transport time, from $T=0.1$ $\mu s$ to $T=20$ $\mu s$, for a fixed perturbation frequency, $\omega=2\pi\times 6$ MHz, in  figure \ref{fig:1}(b). 
Both components periodically reach minimum values for special transport times. 
Moreover, the maxima of the static term remains constant at longer times, while the dynamical term maxima decay and become negligible compared to the static term for very slow shuttling, consistently with Eqs. (\ref{eq:2.39}) and (\ref{eq:2.40}). 
We shall later determine the shortest transport times that make the static contribution dominate.
%
%
%
%
%
\subsection{Envelope functions}
To test the validity of the perturbative treatment, we calculate  $\Delta E_n(T)=E_n(T)-\hbar \Omega(T)(n+1/2)$ by numerically solving Ermakov and Newton equations (\ref{eq:Ermakov}) and (\ref{eq:Newton}) for the auxiliary variables $\rho(t)$ and $q_c(t)$ and inserting them into the equation for the final energy (\ref{eq:energy}). This "exact" result may be compared with the perturbative result, and the differences for the parameters chosen, e.g. in figure 
\ref{fig:1}(b) are hardly noticeable. The main advantage of using the perturbative analysis, instead of numerically solving the differential equations obeyed by $\rho(t)$ and $q_c(t)$, is that we find analytical expressions, which, if lengthy, can be simplified or approximated to get envelope functions. These functions will allow us to find interesting features such as asymptotic behavior at large and small perturbation frequencies or transport times, or to estimate the transport time that makes the static contribution dominate over the dynamical one. 
If the static part dominates, increasing the process time will not improve performance, on average, whereas if the dynamical part dominates, it may be worthwhile to increase the process time.   

We begin with the static contribution to the excitation. Solving the integrals (\ref{eq:19}) and (\ref{eq:20}) for $f(t)=\sin(\omega t)$,  the static term (second line in  (\ref{eq:energy2})) takes the form
\begin{eqnarray}\label{eq:26}
E^{(2)}_{n,stat}(T)&=&\frac{\hbar\Omega_0(2n+1)}{4\left(\omega^2-4\Omega_0^2\right)^2}\Big\{\big[\omega^2\sin{(\omega T)}
\nonumber\\
&-&2\omega\Omega_0\sin{(2\Omega_0 T)}\big]^{\!2}
\!+\!4\omega^2\Omega_0^2\big[\!\cos{(\omega T)}\!-\!\cos{(2\Omega_0 T)}\big]^{2}\Big\}.
\end{eqnarray}
For perturbation frequencies for which $\omega T=k\pi$, i.e.  $\Omega(T)=\Omega_0$ (we will later extend the analysis to arbitrary frequencies), 
\begin{equation}\label{eq:3.5}
E^{(2)}_{n,stat}(T)=\frac{2\hbar\Omega_0(2n+1)}{\left(\omega^2-4\Omega_0^2\right)^2}\omega^2\Omega_0^2\big[1-(-1)^k\cos{(2\Omega_0T)}\big].
\end{equation}
This term vanishes when the condition $(-1)^k\cos{(2\Omega_0T)}=1$ is fulfilled, i.e., when

%
	(i)  $k$ is even $\left(\Leftrightarrow\omega T=2i\pi\right)$ and $2\Omega_0T=2j\pi $ with $i,j\in \mathbb{N}$,
	
	(ii) $k$ is odd $\left(\Leftrightarrow\omega T=(2i'+1)\pi \right)$ and $2\Omega_0T=(2j'+1)\pi $ with $i',j'\in \mathbb{N}$,
	
%
\noindent whereas it is maximum when

	(iii) $k$ is even $\left(\Leftrightarrow\omega T=2i\pi\right)$ and $2\Omega_0T=(2j+1)\pi $ with $i,j\in \mathbb{N}$,
	
	(iv) $k$ is odd $\left(\Leftrightarrow\omega T=(2i+1)\pi \right)$ and $2\Omega_0T=2j'\pi $ with $i',j'\in \mathbb{N}$.

Therefore, by tuning the trap frequency and the transport time appropriately, the final excitation may be minimized. In fact, {\em if $\omega$ is known, one can first choose $T$ and then $\Omega_0$ to fulfill one of the two conditions (i) or (ii) that make the static contribution vanish}. Although in this section we are analyzing a 5th order polynomial protocol, the results for the static contribution are completely general for a sinusoidal perturbation in the trap frequency, as every possible STA trajectory has the same static term. Thus, the choice of $T$ and $\Omega_0$ described to make (\ref{eq:3.5}) vanish holds for any STA protocol.

\begin{figure}[t]
	\begin{center}
\includegraphics[width=0.80\textwidth]{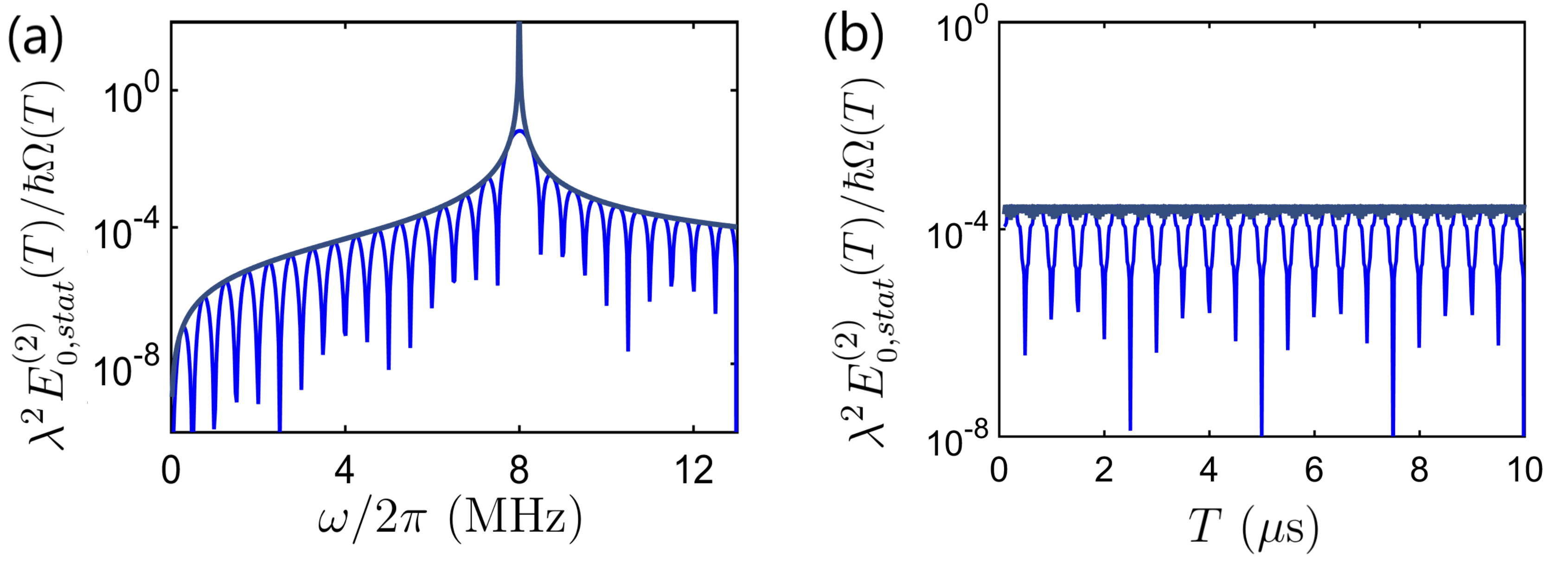}	
		\caption{Log plot of the static component of the excitation (second order term) in units of quanta (light blue line) and its estimated envelope (dark blue line) versus: (a) the sinusoidal perturbation frequency for a transport time of $2\mu$s, and (b) the transport time for a perturbation frequency of $2\pi\times 6$ MHz. }
		\label{fig:5}
		\centering
	\end{center}
\end{figure}

We take as the envelope function of $E^{(2)}_{n,stat}(T)$ the one that goes through all the maxima,
\begin{equation}\label{eq:28}
F_{stat}=\frac{2\hbar\Omega_0(2n+1)}{\left(\omega^2-4\Omega_0^2\right)^2}\omega^2\Omega_0^2\big[1+|\cos{(2\Omega_0T)}|\big].
\end{equation}
The envelope  is  
not valid for very large frequencies ($\omega\gg\Omega_0$) where it decays as $1/\omega^2$, whereas the true static term (\ref{eq:26}) has an oscillating term, proportional to $\sin^2(\omega T)$, that does not decay for large $\omega$.
 In figure \ref{fig:5}(a) we plotted this envelope, together with the static contribution, versus the perturbation frequency for $T=2$ $\mu$s. Even though we only considered a discrete set of frequencies, the envelope is valid for a  continuum of perturbation frequencies.  

In figure \ref{fig:5}(b), we set the perturbation frequency to $\omega=2\pi\times 6$ MHz and let the transport time vary from $0.1$ to $10$ $\mu$s. In this case, the oscillating term in  (\ref{eq:28}) does not add much information, and it could be simply substituted by its maximum value. Again, we should not expect this analysis to work for  $\omega\gg\Omega_0$.

\begin{figure}[t]
	\begin{center}
\includegraphics[width=0.80\textwidth]{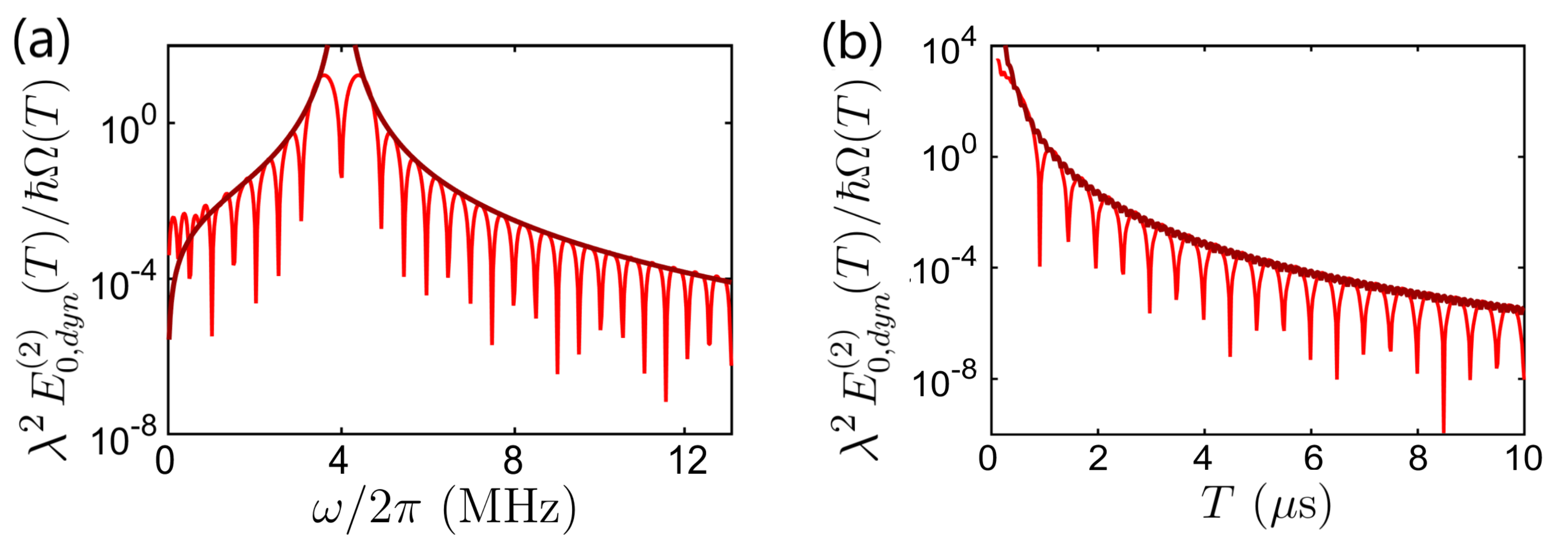}	
		\caption{Log plot of the dynamical component of the excitation in units of quanta (light red line) and its estimated envelope (dark red line) versus: (a) the sinusoidal perturbation frequency for a transport time of $2\mu$s, and (b) the transport time for a perturbation frequency of $2\pi\times 6$ MHz. }
		\centering
		\label{fig:6}
	\end{center}
\end{figure}

While the static contribution has a simple analytical expression for  $f(t)=\sin(\omega t)$,  the dynamical contribution is more complicated. 
Nevertheless, we managed to find an approximate envelope function,
\begin{equation}\label{eq:29}
F_{dyn}=\frac{57600 m\;d^2}{T^6(\Omega_0^2-\omega^2)^4}\omega^2\Omega_0^2\big[1+|\cos{(\Omega_0T)}|\big].
\end{equation}
We compare this function and the dynamical excitation in figure \ref{fig:6} in the perturbation-frequency and the transport-time domains. Despite having neglected many terms,  (\ref{eq:29}) is a good approximation of the true envelope function. The oscillating term, that now has frequency $\Omega_0$ instead of $2\Omega_0$ as in the envelope for the static contribution, could also be substituted by its maximum value in the transport time plot.

In figure \ref{fig:1} the minima of the dynamical contribution approximately coincide with the minima of the static contribution, which we were able to identify analytically. There is an explanation of this feature using the Fourier expressions  (\ref{eq:2.39}) and (\ref{eq:2.40}). The integrals of interest are
\beqa\label{eq:3.8}
&&\int_0^Tdt\sin(\omega t)e^{-2i\Omega_0 t}
=\frac{1}{2i}\!\int_0^T\!\!dt\left(e^{-i(2\Omega_0-\omega) t}-e^{-i(2\Omega_0+\omega) t}\right),
\\
%
%
\label{eq:3.9}
&&\int_0^Tdt\,\ddot{q}_c^{(0)}(t)\sin(\omega t)e^{-i\Omega_0 t}
=\frac{1}{2i}\int_0^Tdt\,\ddot{q}_c^{(0)}(t)\left(e^{-i(\Omega_0-\omega) t}-e^{-i(\Omega_0+\omega) t}\right).
\eeqa
We took $\Omega_0 T=8\pi$, i.e., an even multiple of $\pi$, and so is $2\Omega_0T$. When condition (i) is satisfied, every exponential in Eqs. (\ref{eq:3.8}) and (\ref{eq:3.9}) takes the form $e^{-i2\pi Kt/T}$, where $K$ is an integer. Since the set $\{e^{-i2\pi Kt/T},\,K\in{\mathbb Z}\}$ forms an orthogonal basis for functions with period $T$,  (\ref{eq:3.8}) (the static contribution), vanishes when condition (i) is verified except when $\omega=2\Omega_0$, which corresponds to $K=0$. Something similar happens with (\ref{eq:3.9}). The acceleration of the classical trajectory is an antisymmetric function around $T/2$ (see figure \ref{fig:3.1}(b)) that resembles the function $\sin(2\pi t/T)$. When projected to each of the functions $e^{-i2\pi Kt/T}$, the values $K=\pm1$ will be the most relevant ones. In fact,
\beqa\label{eq:3.10}
\int_0^Tdt\,\ddot{q}_c^{(0)}(t)e^{-i2\pi Kt/T}&=&
\frac{90d}{\pi^2T}\frac{1}{K^3},\hspace{.1cm}\text{if $K\neq0$},
\eeqa
where $\ddot{q}_c^{(0)}(t)$ is deduced from (\ref{eq:24}). $K=0$ gives 0 due to antisymmetry. According to (\ref{eq:3.10}), the most significant projection is achieved for $\lvert\Omega_0-\omega\rvert=2\pi/T$ $(\lvert K\rvert=1)$. Equation (\ref{eq:3.10}) also sets a  $K^{-3}$ scaling for the rest of the projections.
%
%
\subsubsection{Crossing between static and dynamical terms}
One of the motivations to find the envelopes is to estimate at what point the static contribution starts to dominate the excitation energy
and the 
dynamical contribution becomes irrelevant. 

\begin{figure}[]
	\begin{center}
\includegraphics[width=0.98\textwidth]{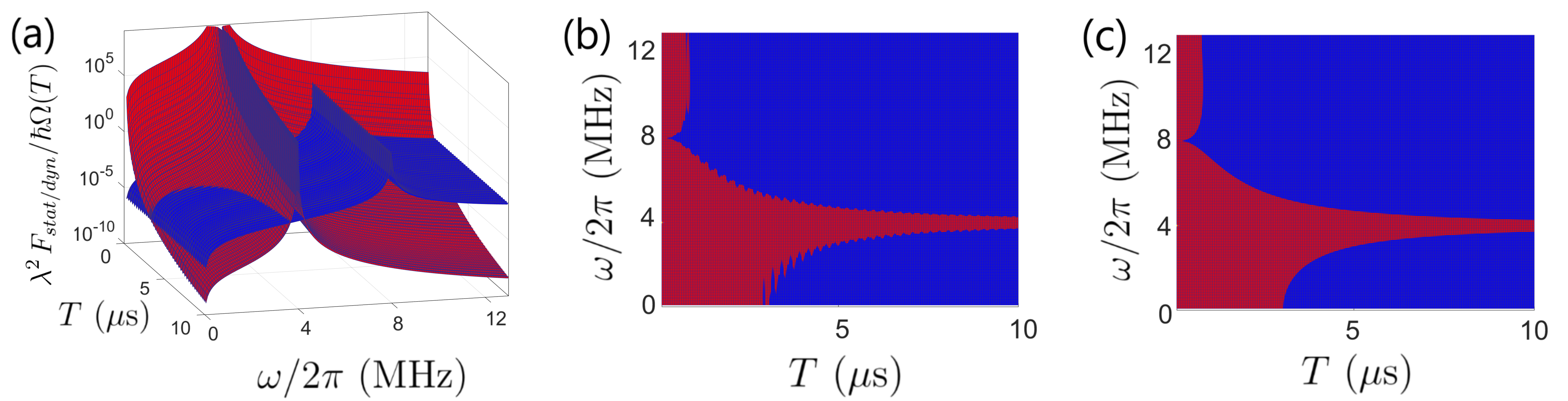}	
		\caption{(a) Log plot of the two envelope functions, static (blue surface) and dynamical (red surface) in Eqs. (\ref{eq:28}) and (\ref{eq:29}), respectively, versus perturbation frequency and the transport time. (b) Top view, showing only the dominant contribution. (c) Top view ignoring the oscillating terms. The parameters $m$, $d$, and $\Omega_0$ are the same as in figure \ref{fig:1}.}
		\centering
		\label{fig:7}
	\end{center}
\end{figure}

In figure \ref{fig:7}(a), we plot the envelopes for the static and dynamical terms as functions of the perturbation frequency and the transport time. 
In figure \ref{fig:7}(b) we present a top view of these two surfaces, showing at each point only the one that dominates. Even if the envelopes are already much simpler than the corresponding contributions to the excitation, the curve defining the crossing points is complicated because of the oscillating terms of  (\ref{eq:28}) and (\ref{eq:29}). In figure \ref{fig:7}(c) we show the envelopes when those oscillating terms are ignored. 
Thus, we find the transport time at which both envelopes cross as a function of $\omega$,
\begin{equation}\label{eq:3.11}
T^*(\omega)=\Bigg\{\frac{28800m\;d^2}{\hbar\Omega_0}\left[\frac{\omega^2-(2\Omega_0)^2}{\left(\omega^2-\Omega_0^2\right)^2}\right]^2\Bigg\}^{1/6}.
\end{equation}
The behavior described by the curve in  (\ref{eq:3.11}) is quite intuitive. {\em Each contribution dominates around its own resonance, $\omega=2\Omega_0$ for the static and $\omega=\Omega_0$ for the dynamical}. When perturbing frequencies, assumed to be given,  are at or near these values  a change of $\Omega_0$ is advisable to avoid excitations.    
\section{Optimal trajectories for an oscillating trap frequency\label{optsec}}
In the previous section we used a polynomial protocol for some given $T$ without trying to optimize performance.  
We will now look for trajectories that minimize the final excitation at second perturbative order when the trap frequency is perturbed sinusoidallly. 
We present several methods that can be applied to find such optimal trap trajectories.

\subsection{Design of the classical trajectory through an auxiliary function (Fourier method)}\label{section:4.1}
For a particle shuttled by a constant-frequency trap the final excitation energy 
is, assuming zero boundary conditions for the trap velocity,  proportional to the Fourier transform of the trap acceleration. 
This  was exploited in a systematic approach 
by Gu\'ery-Odelin and Muga \cite{Guery2014}. The approach makes use of an auxiliary function $g(t)$ to impose  the vanishing of  
\beq\label{eq:2.36}
{\mathcal V}(\Omega_0)=\bigg\lvert\int_0^Tdt\,\ddot{q}_0(t)e^{-i\Omega_0t}\bigg\rvert=0
\eeq
at 
chosen, discrete  values of the trap frequency, see below. 
This method does not use invariants explicitly (even if they are of course implicit) and was devised to transport different species and/or achieve robustness with respect to 
uncertainty or slow changes in the trap frequency, i.e., the trap frequency must be effectively constant throughout each single shuttling process. When condition (\ref{eq:2.36}) is satisfied, and as far as there are no time dependent perturbations affecting the trap parameters, the system ends unexcited. 
Qi {\it et al.} \cite{Muga2021} posed as open questions the applicability or possible generalizations of the method to deal 
with a fast  time dependences of the trap frequency (i.e., noticeable in the scale of $T$), as well as its combination with optimization algorithms.
 In this section we shall first generalize the method in Ref. \cite{Guery2014} to produce  transport without residual excitation 
for a trap-frequency affected by a sinusoidal perturbation. To  optimize the trap trajectories we shall later 
find it more efficient to directly impose conditions of the form (\ref{eq:2.36}) without the need to use an intermediate function $g$. 

To find trajectories that minimize the final excitation we use the Fourier form  (\ref{eq:2.39}) for $f(t)=\sin(\omega t)$. The integral to be minimized is
\beqa\label{eq:36}
\hspace*{-.1cm}{\cal I}(\omega,\Omega_0)\equiv\int_0^T\!dt\,\sin(\omega t)\ddot{q}_c^{(0)}\!(t)e^{-i\Omega_0t}
=\frac{1}{2i}\!\Bigg[\!\!\int_0^T\!\!\!dt\,\ddot{q}_c^{(0)}\!(t)e^{-i(\Omega_0-\omega)t}
\!-\!\!\int_0^T\!\!\!dt\,\ddot{q}_c^{(0)}\!(t)e^{-i(\Omega_0+\omega)t}\Bigg].
\eeqa
Thus, transport  without final excitation at second perturbative order can be achieved by designing a  $q_c^{(0)}(t)$ for which the Fourier transform of its acceleration at $\Omega_0+\omega$ and $\Omega_0-\omega$ takes the same value. One possibility is to cancel it at both  frequencies. Following  \cite{Guery2014} we  introduce an auxiliary function $g(t)$ such that
\beqa\label{eq:4.3}
\ddot{q}_c^{(0)}(t)=\frac{d^4g}{dt^4}(t)+\big[\left(\Omega_0-\omega\right)^2+\left(\Omega_0+\omega\right)^2\big]\frac{d^2g}{dt^2}(t)
+\left(\Omega_0^2-\omega^2\right)^2g(t),
\eeqa
and which obeys the boundary conditions $g(0)=g(T)= \dot{g}(0)=\dot{g}(T)=\ddot{g}(0)=\ddot{g}(T)=g^{(3)}(0)=g^{(3)}(T)=g^{(4)}(0)=g^{(4)}(T)=0$, where dots denote derivatives with respect to time and $g^{(n)}$ is the $n$-th derivative. Equation (\ref{eq:36}) vanishes with such an auxiliary function. We also have to take into account the boundary conditions $q_c^{(0)}(T)=d$ and $\dot{q}_c^{(0)}(T)=0$, which imply that
\beq\label{eq:4.4}
\int_0^Tdt\int_0^{t}dt'g(t')=\frac{d}{\left(\Omega_0^2-\omega^2\right)^2}\;\; \text{and} \; \int_0^Tdt\,g(t)=0.
\eeq
%
\begin{figure}[th]
	\begin{center}
\includegraphics[width=0.85\textwidth]{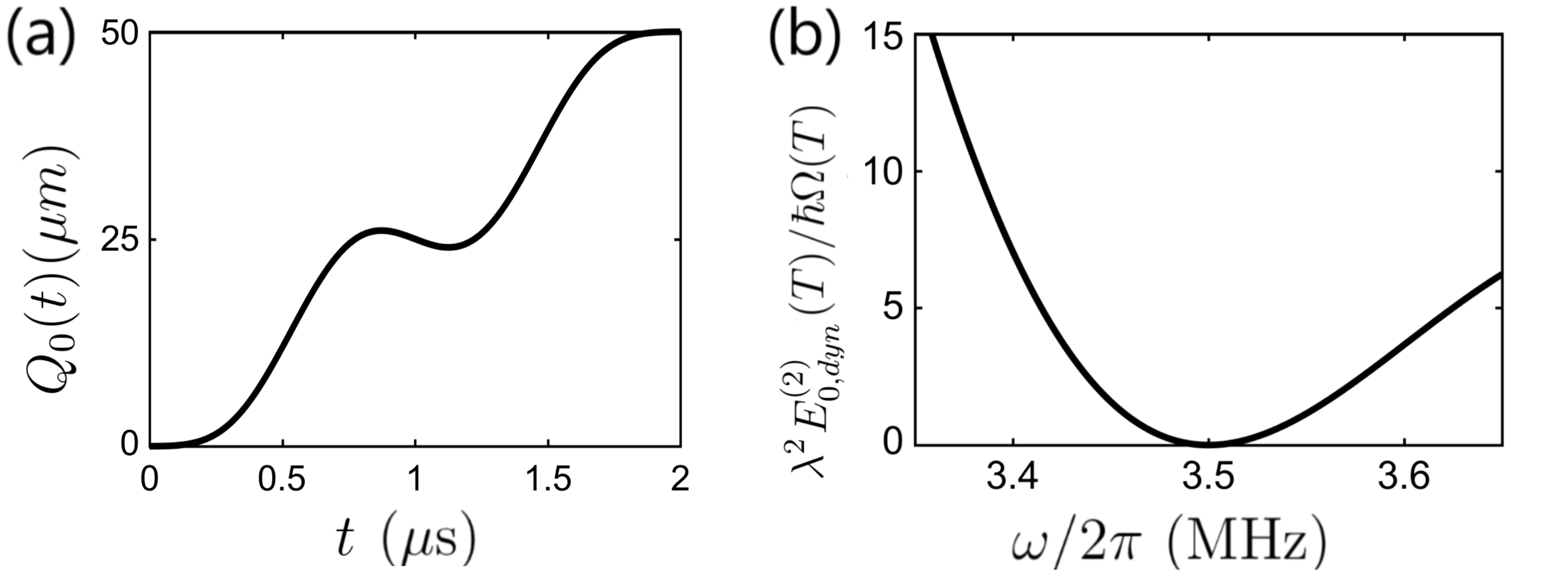}	
		\caption{(a) Trap trajectory  from  (\ref{eq:4.3}) for  $\omega=3.5$ MHz. (b) Final dynamical excitation at second perturbative order and in units of final quanta versus the perturbation frequency for the designed trajectory. The parameters are $\lambda=0.01$, $\Omega_0=2\pi\times4$ MHz, $m=1.455\cdot10^{-25}$ kg ($^{{88}}$Sr$^+$ ion), $d=50\;\mu$m and $T=2\;\mu$s.}
		\centering
		\label{fig:4.1}
	\end{center}
\end{figure}
The auxiliary function $g(t)$ is designed to satisfy its boundary conditions and  (\ref{eq:4.4}). From  $g(t)$, $\ddot{q}_c^{(0)}(t)$ is deduced via  (\ref{eq:4.3}). Then, we integrate this expression twice to get the classical trajectory. Let us consider, similarly to \cite{Guery2014}, the simple form
\beq\label{eq:4.5}
g(t)={\cal N}\left(\frac{t}{T}\right)^5\left(\frac{t}{T}-1\right)^5\left(\frac{t}{T}-\frac{1}{2}\right),
\eeq
where ${\cal N}$ is a normalization factor that has to be deduced from the first condition of  (\ref{eq:4.4}). The second and third factors in  (\ref{eq:4.5}) guarantee the boundary conditions at initial and final times, while the fourth one provides the odd symmetry to satisfy the second condition in  (\ref{eq:4.4}). In figure \ref{fig:4.1} we show the results found using this method for  $\omega=3.5$ MHz. We have plotted  the trap trajectory $Q_0(t)$ and the final excitation (in quanta units) versus the perturbation frequency $\omega$ around $3.5$ MHz. The rest of the parameters are the same as the ones used in the previous section. 
We observe a vanishing excitation at the perturbation frequency used to design  the trajectory.

 The procedure to 
make the protocol robust for a range of trap frequencies in \cite{Guery2014} can be adapted to our problem. Suppose that there are multiple perturbation frequencies $\omega_1$, $\omega_2$,..., $\omega_p$ affecting the shuttling operation. In order for the protocol to provide an excitation-free final state (at second perturbative order of the excitation), the classical acceleration may be written  as
$$
\ddot{q}_c^{(0)}(t)\!=\!P_0g^{(4p)}+P_1g^{(4p-2)}+\cdots+ P_jg^{(4p-2j)}+\cdots+P_{2p}g(t),
$$
where    
\beqa
P_0&=&1,\hspace{1cm} P_1=\sum_{i=1}^p\sum_{\sigma_i=\{+,-\}} (\Delta_i^{\sigma_i})^2,
\nonumber\\
P_2&=&\sum_{i<j}\sum_{\sigma_{i,j}=\{+,-\}} (\Delta_i^{\sigma_i})^2(\Delta_j^{\sigma_j})^2, \dots\nonumber\\
P_{2p}&=& (\Delta_1^+)^2(\Delta_1^-)^2\cdots (\Delta_p^+)^2(\Delta_p^-)^2,\nonumber
\eeqa
and
%
$\Delta_i^\pm=\Omega_0\pm\omega_i. 
$
%
Now, the function $g(t)$ should have $8p+2$ vanishing boundary conditions,
$$
g(0)\!=\!g(T)\!=\!\dot{g}(0)\!=\!\dot{g}(T)\!=\!\cdots\!=\!g^{(4p)}(0)\!=\!g^{(4p)}
(T)=0.
$$
If the perturbation frequencies are distributed in a continuous region, the robustness is achieved by choosing the $p$ frequencies close enough in that region, flattening the excitation in a window of frequencies.
\subsection{Fourier ansatz for the classical acceleration}\label{section:4.2}
With the method described in the previous subsection, the number of boundary conditions imposed on $g(t)$ escalates considerably to increase  robustness. To solve this problem and avoid the use of an an intermediate function $g(t)$, we now choose a different, direct ansatz,
\beq\label{eq:4.8}
\ddot{q}_c^{(0)}(t)=\sum_{j=1}^N a_j \sin(j\pi t/T).
\eeq
The boundary conditions $\ddot{q}_c^{(0)}(0)=\ddot{q}_c^{(0)}(T)=0$ are automatically satisfied, and the number of terms $N$ will depend on the number of constrains imposed on $q_c^{(0)}(t)$ and its derivatives, whereas the $\{a_j\}$ will be determined from them. We integrate  (\ref{eq:4.8}) to specify the classical trajectory and velocity,
\beqa
\hspace*{-.4cm}q_c^{(0)}(t)&=&\int_0^tdt'\int_0^{t'}dt''\ddot{q}_c^{(0)}(t'')
=\sum_{j=1}^N a_j\frac{T}{(j\pi)^2} \left[j\pi t-T\sin(j\pi t/T)\right],
%
\\
\hspace*{-.4cm}\dot{q}_c^{(0)}(t)&=&\int_0^tdt'\ddot{q}_c^{(0)}(t')=\sum_{j=1}^N \!a_j\frac{T}{j\pi}\!\left[1\!-\! \cos(j\pi t/T)\right]\!.
\eeqa
It can be checked that the initial conditions $q_c^{(0)}(0)=0$ and $\dot{q}_c^{(0)}(0)=0$ are fulfilled. The final time 
boundary conditions  $q_c^{(0)}(T)=d$ and $\dot{q}_c^{(0)}(T)=0$, lead to two conditions on the coefficients $\{a_j\}$,
\beq\label{eq:4.11}
\sum_{j=1}^N a_j \frac{T^2}{j\pi}=d,\;\;\;\;\;\;
%
\sum_{j=1}^N \frac{a_j}{j} \left[1-(-1)^j\right]=0.
\eeq 
To cancel the final excitation, the $\{a_j\}$ must also verify
\beqa\label{eq:4.13}
{\cal I}(\omega,\Omega_0)&=&\sum_{j=1}^N a_j {\cal I}_j(\omega,\Omega_0)=0, 
\\
{\cal I}_j(\omega,\Omega_0)&\equiv&\int_0^Tdt\,\sin(\omega t)\sin(j\pi t/T)e^{-i\Omega_0t}\nonumber\\
&=&\frac{T}{2}\Bigg\{\frac{i\Omega_0 T}{(j\pi-\omega T)^2-(\Omega_0 T)^2}-\frac{i\Omega_0 T}{(j\pi+\omega T)^2-(\Omega_0 T)^2}\nonumber\\
&+&e^{-i\Omega_0 T}\bigg[\frac{\left(j\pi-\omega T\right)\sin(j\pi-\omega T)-i\Omega_0 T \cos(j\pi-\omega T)}{(j\pi-\omega T)^2-(\Omega_0 T)^2}\nonumber\\
&-&\frac{\left(j\pi+\omega T\right)\sin(j\pi+\omega T)-i\Omega_0 T \cos(j\pi+\omega T)}{(j\pi+\omega T)^2-(\Omega_0 T)^2}\bigg]\Bigg\}.\nonumber
\end{eqnarray}
which are in fact two conditions, since  the real and the imaginary parts have to be canceled. Together with conditions (\ref{eq:4.11}), 
there are 4 equations for the coefficients $\{a_j\}$, and thus, at least $N=4$ terms are needed to define the classical trajectory.
\begin{figure}
	\begin{center}
\includegraphics[width=0.98\textwidth]{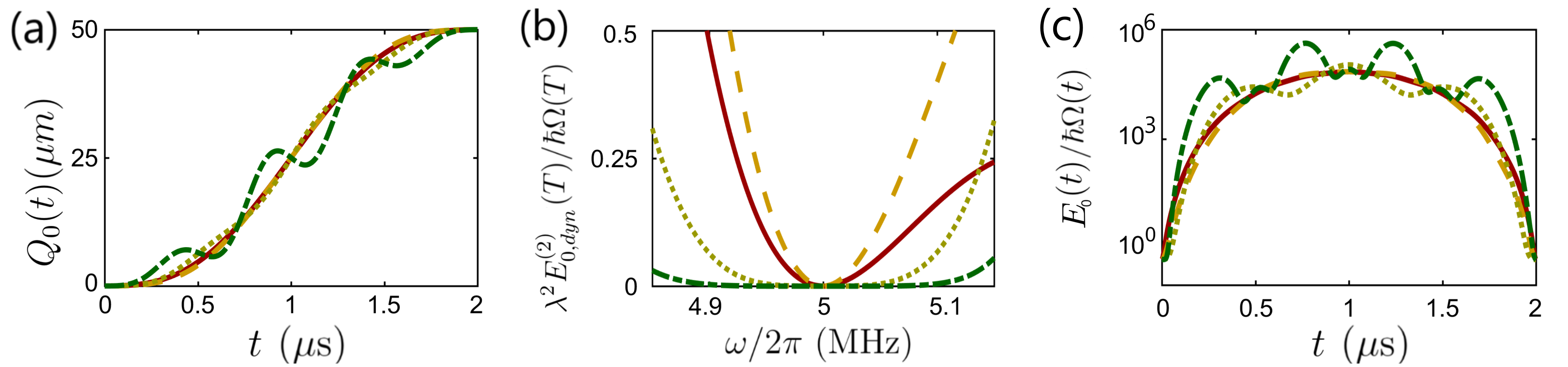}			
\caption{(a) Transport function for a protocol that cancels the integral from  (\ref{eq:36}) (red solid line), and up to its first (yellow dashed line), second (light green dotted line) and third (dark green dash-dotted line) derivatives with respect to the perturbation frequency when  $\Omega_0=2\pi\times 4$ MHz and $\omega=2\pi\times5$ MHz. The parameters are: $^{88}$Sr$^+$ ion, $d=50$ $\mu$m,  $T=2$ $\mu$s, and $\lambda=0.01$. (b) Dynamical component of the final excitation energy in units of final quanta in  each of the protocols versus $\omega$.
(c) Transient energy  (in units of quanta)  during the transport for the protocols shown in  figures (a) and (b) (same line and color code).
\label{fig:8}}
	\end{center}
\end{figure}
More terms can be added to increase robustness. For instance, 
to have excitation-free final states for a range of perturbation frequencies, we may impose the cancellation of the derivatives of ${\cal I}(\omega,\Omega_0)$ from  (\ref{eq:36}) with respect to $\omega$. For every derivative nullified, we have to add at least two terms in (\ref{eq:4.8}) for the system of equations relating the $\{a_j\}$ not to be overdetermined. In figure \ref{fig:8} we compare different transport protocols: $N=4$ with no restriction on the derivatives; $N=6$ with cancellation of first derivative with respect to $\omega$; $N=8$ with cancellation of the first two derivatives; and $N=10$ with cancellation of first three derivatives. The coefficients $\{a_j\}$ are uniquely determined. The parameters are the same as the ones used in  figure \ref{fig:1} or figure \ref{fig:4.1}. We clearly observe in figure \ref{fig:8}(b) an increase of the robustness against the perturbation frequencies when the number of canceled derivatives increases. A price to pay is a more oscillatory behavior in the trap trajectory $Q_0(t)$, see figure \ref{fig:8}(a) (let us recall that the trap trajectory is related to the classical trajectory $q_c(t)$ through  (\ref{eq:14})), which may involve larger transient energies, see figure \ref{fig:8}(c). 
\begin{figure}[]
	\begin{center}
\includegraphics[width=0.80\textwidth]{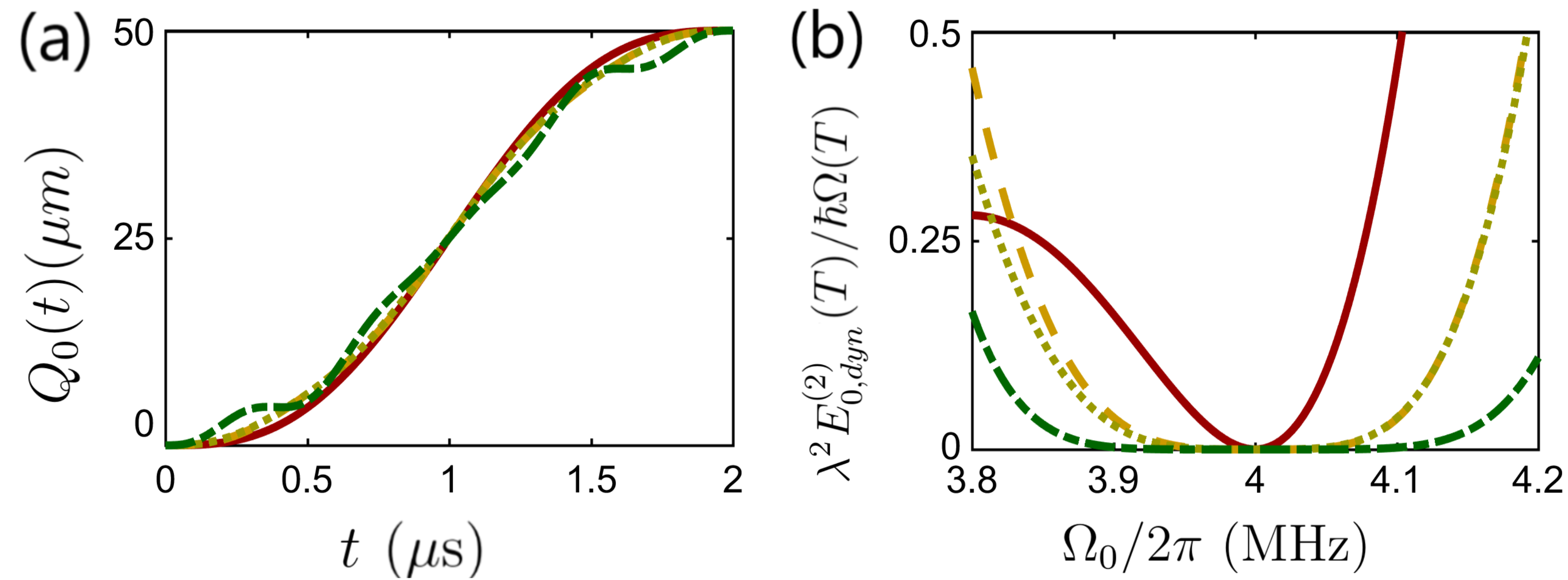}	
		\caption{(a) Transport function for a protocol that cancels the integral in  (\ref{eq:36}) (red solid line), and up to its first (yellow dashed line), second (light green dotted line) and third (dark green dash-dotted line) derivatives with respect to $\Omega_0$  when $\Omega_0=2\pi\times 4$ MHz and  $\omega=2\pi\times5$ MHz. The parameters are: $^{88}$Sr$^+$ ion,  $d=50$ $\mu$m, $T= 2$ $\mu$s, and $\lambda=0.01$. (b) Dynamical component of the final excitation energy in units of final quanta  versus the trap frequency.}
		\centering
		\label{fig:9}
	\end{center}
\end{figure}
Similarly, the concept of robustness can be extended to other errors. For instance, suppose that, aside from the sinusoidal perturbation, the central trap frequency $\Omega_0$ takes different values over multiple runs of a transport experiment. 
Robustness with respect to these deviations can be achieved by imposing the cancellation of the derivatives of ${\cal I}(\omega,\Omega_0$) with respect to $\Omega_0$. In figure \ref{fig:9} we show again 4 different protocols: $N=4$ with no restriction on the derivatives; $N=6$ with cancellation of first derivative; $N=8$ with cancellation of the first two derivatives; and $N=10$ with cancellation of first three derivatives. As in the previous case, the coefficients $\{a_j\}$ are uniquely determined. Now, the protocols increase the robustness against variations of $\Omega_0$ when the number of nullified derivatives increases.
These  ideas can be combined, simultaneously canceling derivatives with respect to $\omega$ and $\Omega_0$ and making the protocol robust against variations of both of them. 
\subsection{Comparison between the auxiliary function and Fourier ansatz methods}
The methods  in subsections \ref{section:4.1} and \ref{section:4.2} lead to trajectories that leave the ion in its final position without final dynamical excitation up to second perturbation order. Both rely on nullifying the integral   (\ref{eq:36}). However, the Fourier ansatz  is more straightforward, since it does not involve additional steps to design an auxiliary function. The auxiliary function method forces two integrals to vanish for each perturbation frequency (see  (\ref{eq:36})) instead of their sum. 

\begin{figure}[]
	\begin{center}
\includegraphics[width=0.89\textwidth]{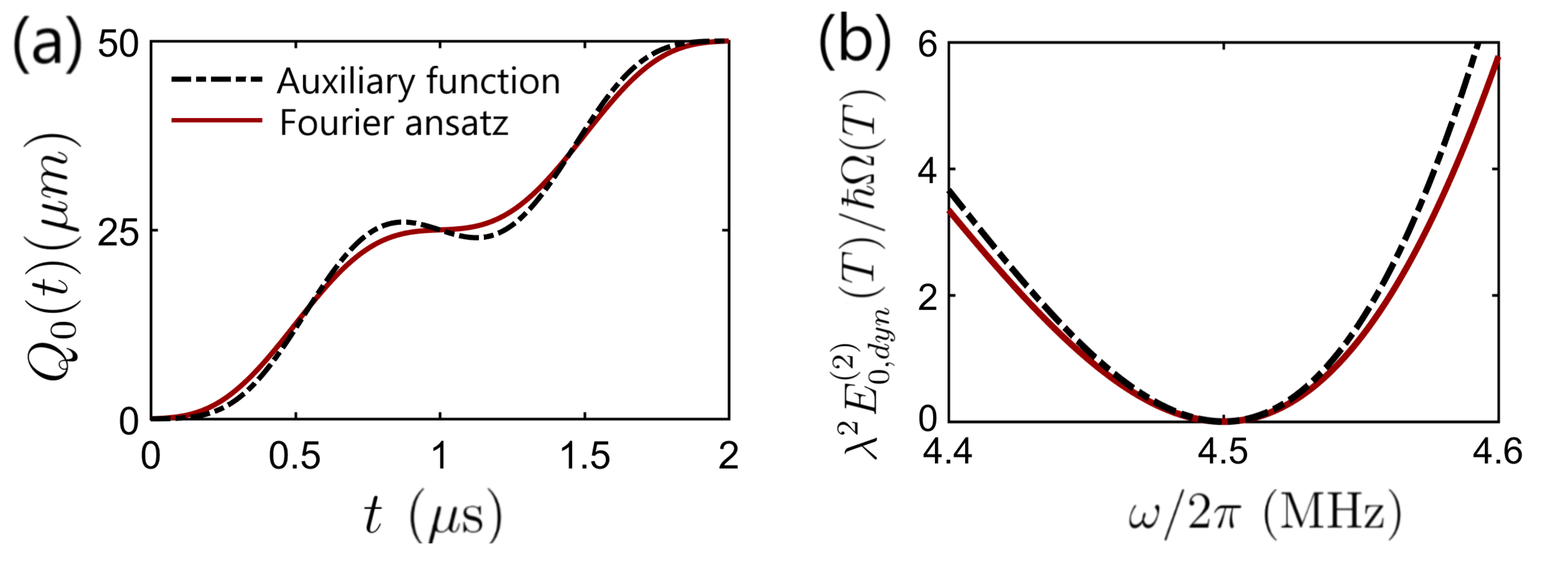}	
		\caption{Comparison between the method based on an auxiliary function, see  \ref{section:4.1}, and the method based on a Fourier ansatz for $\ddot{q}_c^{(0)}(t)$. (a) Trap trajectories with zero final excitation (up to second perturbation order) 
		for $\omega=4.5$ MHz. (b) Final excitation versus the perturbation frequency for the trajectories in the left figure. The rest of parameters are the same as in figure \ref{fig:4.1}.}
		\centering
		\label{fig:9.1}
	\end{center}
\end{figure}

In figure \ref{fig:9.1} we have compared the two methods by finding the trajectories for which the second perturbation dynamical excitation is zero for a fixed perturbation frequency $\omega_{target}=4.5$ MHz, using the same parameters in figure \ref{fig:4.1}. Both methods are used without applying additional conditions to flatten the excitation, that is, in their most basic forms (cancellation of up to the 4th derivative of g(t) at its bounds and $N=4$ terms in the Fourier ansatz). The final dynamical excitation curve versus an $\omega$ actually applied is lower with the trajectory found with the Fourier ansatz for  $\omega_{target}$, see figure \ref{fig:9.1}(b). 
The trajectory given by the Fourier ansatz is also smoother, avoiding significant accelerations during the shuttling. 

Therefore, {\em the Fourier ansatz method presents advantages in simplicity and effectiveness over the method that uses an auxiliary function}. In the following section, we apply the Fourier sum ansatz in combination with a genetic algorithm. 
\subsection{Genetic algorithms \label{section:4.3}}
The method  in  subsection \ref{section:4.2} can be generalized for further flexibility by including more terms in  (\ref{eq:4.8}) and applying more conditions. 
For example,  trap trajectories that do not exceed the range from the initial to the final position, i.e., $0<Q_0(t)<d$, are highly preferable. Although this condition is fulfilled by the protocols in figures \ref{fig:8} and \ref{fig:9}, it is not generally satisfied. Short transport times, perturbation frequencies close to the trap frequency, and cancellation of too many derivatives may lead to trajectories that go beyond these limits.  

A solution is to leave the system of equations for the $\{a_j\}$ underdetermined by letting $N$ be greater than the number of conditions. Then the coefficients may be chosen by minimizing  a given cost function. For instance, to limit the trajectory inside its boundaries $[0,d]$, the cost function can be
\beq\label{eq:cf}
\hspace*{-.2cm}f\!=\!\!\int_0^T\!\!dt\,F[Q_0(t)],\,\,\;\;\;\;\;\;\;
%
F[Q_0(t)]\! =\! 
\begin{cases}
	Q_0(t)\!-\!d, &Q_0>d\\
	0, &0\!\le\! Q_0\!\le\! d\\
	-Q_0(t),\; &Q_0<0, \\ 
\end{cases}
\eeq
with the least possible number of terms $N$ defining $Q_0$.

Genetic algorithms are versatile optimization methods methods where a population of individuals evolve through selection, crossover and mutation towards better solutions, inspired by natural selection \cite{Goldberg1989}. 
\begin{figure}[]
	\begin{center}
\includegraphics[width=0.48\textwidth]{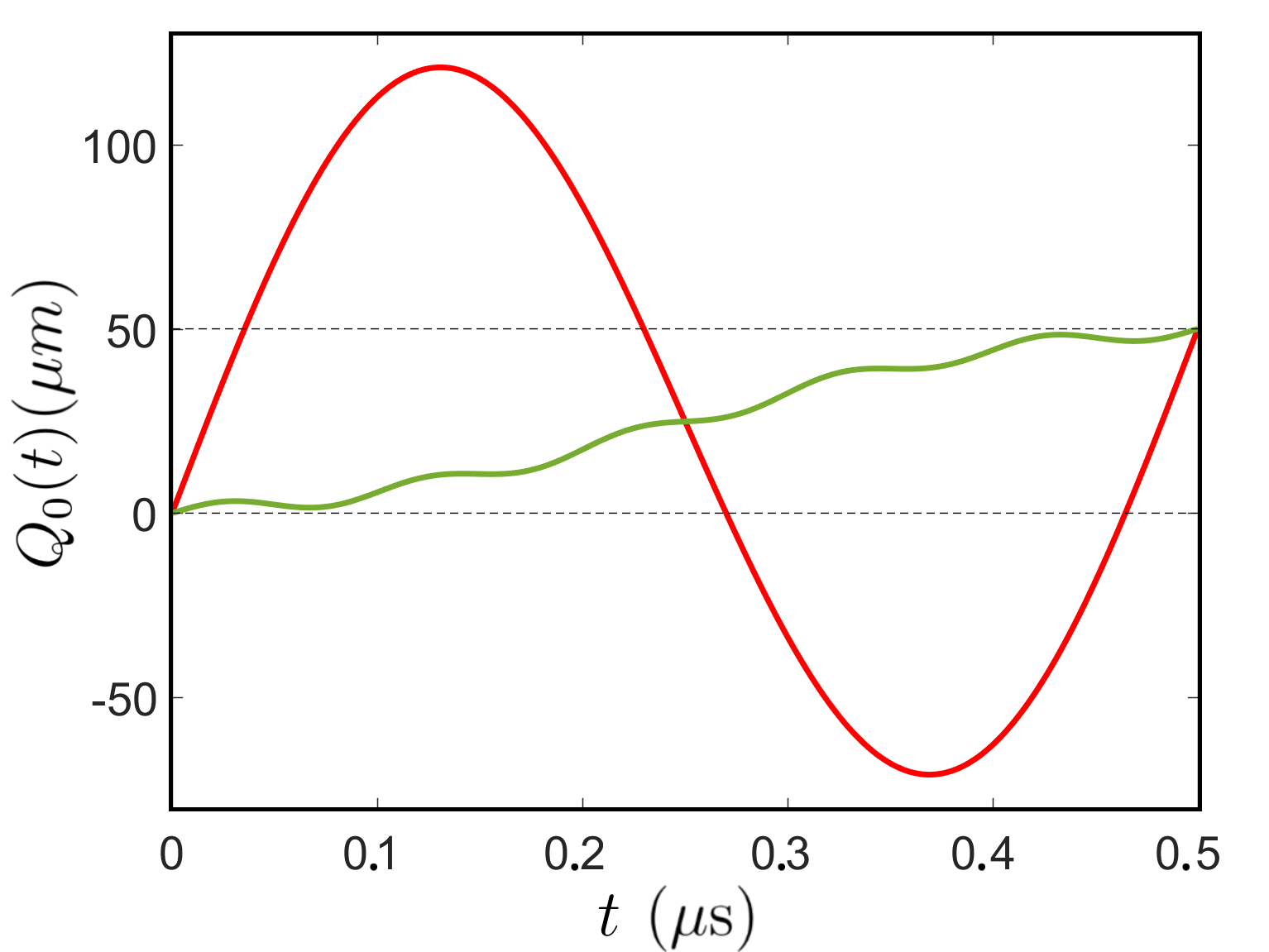}	
		\caption{Trap trajectories that cancel the integral from  (\ref{eq:4.13}) for a perturbation frequency $\omega=2\pi\times5$ MHz, a trap frequency $\Omega_0=2\pi\times4$ MHz, and a transport time $T=0.5$ $\mu$s. The red curve is for $N=4$ terms in the ansatz (\ref{eq:4.8}), while the green curve is for $N=10$ and letting the genetic algorithm minimize condition (\ref{eq:cf}).}
		\centering
		\label{fig:10}
	\end{center}
\end{figure}
In our problem, each individual is a set of $N$ coefficients $\{a_j\}$ such that conditions (\ref{eq:4.11})  and (\ref{eq:4.13}) are verified. The algorithm stops whenever the result of integral (\ref{eq:cf}) is zero, or when too many generations give the same value for the integral, meaning that the algorithm has fallen into a local minimum and mutations are not enough to jump to a better minimum. In figure \ref{fig:10} we compare the trap trajectory that satisfies the aforementioned conditions  found for $N=4$, which is the unique solution, since we have 3 real + 1 imaginary conditions (red line), with a trap trajectory found by the genetic algorithm for $N=10$ (green line). The short transport time ($T=0.5$ $\mu$s) makes the first protocol to exceed the interval $[0,d]$,  while the  solution by the genetic algorithm stays inside $[0,d]$.
\subsection{Optimal Control Theory\label{section:4.4}}
Invariant-based inverse engineering may be combined with optimal control theory via Pontryagin's principle \cite{Pontryagin}, see 
Chen {\it et al} \cite{Chen2011} and more examples and references in Gu\'ery-Odelin {\it et al.}  \cite{Guery2019}. 
In this section, we will apply the OCT formalism 
to minimize the transient potential energy. Let us define first the state variables
\beqa
x_1(t)&=&q_c^{(0)}(t),\h x_2(t)=\dot{q}_c^{(0)}(t),
\nonumber\\ 
x_3(t)&=&q_c^{(1)}(t),\h x_4(t)=\dot{q}_c^{(1)}(t)
\eeqa
and (scalar) control function
\beq
u(t)=q_c^{(0)}(t)-Q_0(t).
\eeq
Equations (\ref{eq:14}) and (\ref{eq:18}) give a system of equations with the form $\mathbf{\dot{x}}=\mathbf{f}\big[\mathbf{x}(t),u(t)\big]$, that is
\beqa\label{eq:4.15}
\dot{x}_1(t)&=&x_2(t),\;\;
\dot{x}_2(t)=-\Omega_0^2u(t),\;\;
\dot{x}_3(t)=x_4(t),\nonumber\\
\dot{x}_4(t)&=&-\Omega_0^2x_3(t)-2\Omega_0^2\sin(\omega t)u(t).
\eeqa
Our optimal control problem is to minimize some cost function. We choose to minimize the average dynamical term of the potential energy,
\beq\label{eq:4.20}
\overline{E}_{P,dyn}=\frac{1}{T}\int_0^Tdt\frac{m\Omega^2(t)}{2}\left[q_c(t)-Q_0(t)\right]^2,
\eeq
which, assuming small $\lambda$, can be approximated by
\beq\label{eq:4.16}
\overline{E}_{P,dyn}\approx\frac{1}{T}\frac{m\Omega_0^2}{2}\int_0^Tdt\left[u(t)\right]^2.
\eeq
Equation (\ref{eq:4.16}) uses only the zeroth order approximation for the energy. We shall later demonstrate that this  order is enough to account for the transient energy. The reason is that, unlike the final energy, the zeroth order of the energy takes a nonzero value during the transport, and therefore higher perturbative orders are negligible in comparison.
Thus, from  (\ref{eq:4.16}) the cost function is
\beq\label{eq:4.21}
J(u)=\int_0^Tdt\,\left[u(t)\right]^2.
\eeq
For an excitationless transport, the boundary conditions that have to be satisfied are (i)  (\ref{eq:6}) and (\ref{eq:15}) for $q_c^{(0)}$ and $\dot{q}_c^{(0)}$, and (ii) the cancellation of $q_c^{(1)}$ and $\dot{q}_c^{(1)}$ at the endpoints, to make the first line in  (\ref{eq:energy2}) (the dynamical excitation) vanish at final time. This implies that the dynamical system starts and ends at 
\beq\label{eq:4.24}
\mathbf{x}(0)=\begin{pmatrix} x_1(0)\\ x_2(0)\\ x_3(0)\\x_4(0) \end{pmatrix}=\begin{pmatrix} 0\\ 0\\ 0 \\ 0 \end{pmatrix},\;
%
%
\mathbf{x}(T)=\begin{pmatrix} x_1(T)\\ x_2(T)\\ x_3(T)\\x_4(T) \end{pmatrix}=\begin{pmatrix} d\\ 0\\ 0 \\ 0 \end{pmatrix},
\eeq
The additional conditions $Q_0(0)=0$ and $Q_0(T)=d$ are translated to the control parameter as $u(0)=u(T)=0$. At these points, jumps of the optimal control will be required to match these boundary conditions. 

To minimize the cost function (\ref{eq:4.21}), we apply Pontryagin’s maximal principle. The control Hamiltonian is
\beq
H_c=p_1x_2-p_2\Omega_0^2u+p_3x_4-p_4x_3\Omega_0^2-2p_4\Omega_0^2\sin(\omega t)u-p_0u^2,
\eeq
where $p_0$ is a normalization constant greater than 0, and $\{p_1,\,p_2,\,p_3,\,p_4\}$ are the costates (time dependences of the state, costate and control variables have been dropped to simplify the notation). Pontryagin’s maximal principle states that for the dynamical system $\mathbf{\dot{x}}= \mathbf{f}(\mathbf{x}(t), u(t))$, the coordinates of the extremal vector $\mathbf{x}(t)$ and of the corresponding adjoint sate $\mathbf{p}(t)$ fulfill $\mathbf{\dot{x}}=\partial{H_c}/\partial{\mathbf{p}}$ and $\mathbf{\dot{p}}=-\partial{H_c}/\partial{\mathbf{x}}$, which gives the four costate equations
\beq
\dot{p}_1(t)=0,\;\;\;\;\;\;
\dot{p}_2(t)=-p_1(t),\;\;\;\;\;\;
\dot{p}_3(t)=\Omega_0^2p_4(t),\;\;\;\;\;\;
\dot{p}_4(t)=-p_3(t).\label{eq:4.26}
\eeq
According to the maximum principle, the control $u(t)$ maximizes the control Hamiltonian at each time. For simplicity, we choose $p_0=\Omega_0^2/2$, so that the minimal condition of the control Hamiltonian $\partial{H_c}/\partial{u}=0$ gives
\beq
u(t)=-\left[p_2(t)+2\sin(\omega t)p_4(t)\right],
\eeq
whereas, from (\ref{eq:4.26}), we get
\beqa
p_1(t)&=&-c_1,\;\;\,
p_2(t)=c_1t+c_2,\\
p_3(t)&=&c_3\Omega_0\sin\left(\Omega_0 t\right)-c_4\Omega_0\cos\left(\Omega_0 t\right),\\
p_4(t)&=&c_3\cos\left(\Omega_0 t\right)+c_4\sin\left(\Omega_0 t\right),
\eeqa
Where $c_1,...,c_4$ are constants that will eventually be determined from the boundary conditions on $\mathbf{x}(t)$. Substituting $p_2$ and $p_4$ into $u(t)$, and then inserting $u(t)$ in the system  (\ref{eq:4.15}), we find 
explicit but somewhat lengthy expressions for $x_1$, $x_2$, $x_3$, and $x_4$, not shown here. 
Finally, the trap trajectory is determined as
$$
Q_0(t)=  
\begin{cases}
	0, & t\le0\\
	x_1(t)-u(t),    & 0<t<T\\
	d,& t\ge T
\end{cases}.
$$

\begin{figure}[]
	\begin{center}
\includegraphics[width=0.98\textwidth]{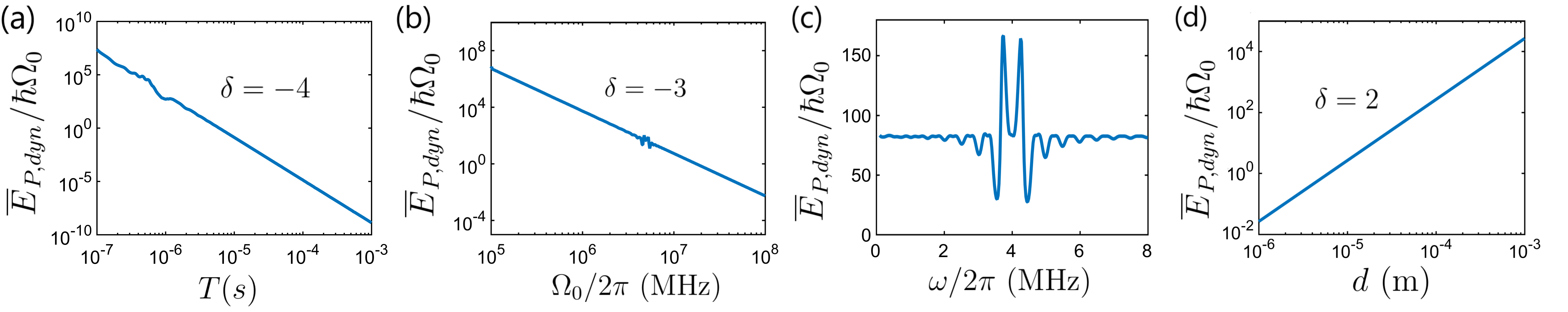}	
		\caption{Time average of the dynamical contribution to the potential energy using  (\ref{eq:4.20}) including the first perturbation order of $q_c(t)$ during the transport of a $^{88}$Sr$^+$ ion  versus (a) the transport time $T$, (b) the trap frequency $\Omega_0$, (c) the perturbation frequency $\omega$, and (d) the shuttling distance $d$. In each figure, the parameters that do not vary are kept at $T=2$ $\mu$s, $\Omega_0=2\pi\times4$ MHz, $\omega=2\pi\times5$ MHz and $d=50$ $\mu$m. The $\delta$ is the asymptotic exponent of each of the parameters.}
		\centering
		\label{fig:11}
	\end{center}
\end{figure}
The discontinuities of $Q_0(t)$ at $t=0$ and $t=T$ may prevent these trajectories to be experimentally feasible, but they provide, in any case, a lower bound for the time average of $E_P$. The values of the coefficients $c_1-c_8$ depend mainly on  $T$,  $\Omega_0$, $\omega$ and $d$. In figure \ref{fig:11}, we show the average dynamical potential energy as a function of each of these 4 parameters, while keeping the rest fixed (see caption for further details). Although for the optimal control problem we have only considered the zeroth order of the transient energy, in the calculations for this figure we have included the first order perturbative term of $q_c(t)$. The results are indistinguishable to those in which they are not considered. The asymptotic behavior, away from oscillations around $\omega=\Omega_0$, is given by a power law of $T$, $\Omega_0$ and $d$,
\beq
\overline{E}_{P,dyn}\varpropto\frac{d^2}{T^4\Omega_0^2},
\eeq
in agreement with the lower bound found for the average potential energy in the unperturbed case \cite{Torrontegui2011}. The perturbation does not change the asymptotic behavior of the mean potential energy. This is also show in figure \ref{fig:11}(c), where the curve flattens far from $\omega=\Omega_0$. 

\section{Conclusion\label{consec}}
In this work, we studied the effect of small perturbations in some of the trap parameters in shortcuts-to-adiabaticity (STA) shuttling protocols of an ion driven by a harmonic trap, with emphasis in sinusoidal perturbations or their combinations. We have also found robust protocols with respect to these perturbations.   

We have applied the invariant-based inverse engineering formalism, combined with a perturbative treatment, to find expressions of the final excitation when the perturbation affects the trap frequency or the trap trajectory, identifying static and dynamical terms (independent and dependent, respectively,  on the ideal STA trajectory). We have also found for these terms simple Fourier integral forms.  Quite generally the static contribution is worse for perturbed trajectories than for perturbed frequencies which suggests to  put the emphasis in  implementing the trajectory faithfully in inversion subroutines from the ideal trajectories to the implemented electrode voltages in multisegmented Paul traps.     

We have thoroughly analyzed the basic 5th order polynomial STA protocol to shuttle a particle for a distance $d$ in a time $T$ for a sinusoidally perturbed trap frequency.  (The analysis for the static contribution is generic and valid for  any STA protocol.) We could determine points with no final (static and dynamical) excitation when the perturbation frequency is known. We also found conditions, in particular minimal times, for the static contribution to dominate.  

Finally we have presented several  techniques to optimize the driving for sinusoidally perturbed trap frequencies with respect to  
final energy for a span of perturbation frequencies;  trajectory domain;  or average transient energy. These techniques are flexible and complementary, they could be applied to other objectives as well. In particular the same approaches could be applied for 
perturbations in the trajectory. 
Both methods described in  \ref{optsec}\ref{section:4.1} (auxiliary function) and \ref{optsec}\ref{section:4.2} (Fourier ansatz) to design the classical acceleration increase  robustness by widening the window of perturbation frequencies for quiet transport. We found better results with the Fourier ansatz method when comparing the most basic approaches, but notice that the auxiliary function method admits unexplored generalizations with different auxiliary functions. The Fourier ansatz method can be easily combined with optimization algorithms, such as genetic algorithms, as  in  \ref{optsec}\ref{section:4.3}. Although we have focused on limiting the trap trajectory inside the range $[0,d]$, genetic algorithms can be used for a broad span of optimizations (bounded velocity, minimal peak transient energy,...). A problem with genetic algorithms is that there is no guarantee that the solution found is the global minimum, and many runs of the algorithm could be  needed to find an optimal trajectory. On the other hand, optimal control theory offers the tools to find analytical bounds and asymptotic behavior, even if the solutions may contain discontinuities that make them hard to implement experimentally.

\vskip6pt

\enlargethispage{20pt}



\aucontribute{HE carried out the calculations and drafted the manuscript. XJL worked on the perturbative analysis. JE provided support on the 
numerical work. JGM designed the study. All authors read and approved the manuscript.}

\competing{The authors declare that they have no competing interests.}

\funding{This work was supported by the Basque Country Government (Grant No. IT986-16),  by the Spanish Ministry of Science and Innovation through projects PGC2018-101355-B-I00 and PGC2018-095113-B-I00 (MCIU/AEI/FEDER,UE), and by the 
Natural Science Foundation of Henan Province (Grant No. 212300410238).
}

\ack{We thank A. Ruschhaupt, D. Gu\'ery-Odelin,  E. Torrontegui, J. Chiaverini, and L. Chi  for many discussions.}


\bibliography{BibliographyTFM}{}

\providecommand{\href}[2]{#2}\begingroup\raggedright\begin{thebibliography}{10}

\bibitem{Torrontegui2013}
E.~Torrontegui, S.~Ib{\'{a}}{\~{n}}ez, S.~Mart{\'{i}}nez-Garaot, M.~Modugno,
  A.~del Campo, D.~Gu{\'{e}}ry-Odelin, A.~Ruschhaupt, X.~Chen, and J.~G. Muga,
  ``{\em {Shortcuts to Adiabaticity}}'',
  \href{http://dx.doi.org/10.1016/B978-0-12-408090-4.00002-5}{Adv. At. Mol.
  Opt. Phys. {\bfseries 62},  117--169 (2013)}.

\bibitem{Guery2019}
D.~Gu\'ery-Odelin, A.~Ruschhaupt, A.~Kiely, E.~Torrontegui,
  S.~Mart\'{\i}nez-Garaot, and J.~G. Muga, ``{\em Shortcuts to adiabaticity:
  Concepts, methods, and applications}'',
  \href{http://dx.doi.org/10.1103/RevModPhys.91.045001}{Rev. Mod. Phys.
  {\bfseries 91},  045001 (2019)}.

\bibitem{Lu2020}
X.-J. Lu, A.~Ruschhaupt, S.~Martínez-Garaot, and J.~G. Muga, ``{\em Noise
  Sensitivities for an Atom Shuttled by a Moving Optical Lattice via Shortcuts
  to Adiabaticity}'', \href{http://dx.doi.org/10.3390/e22030262}{Entropy
  {\bfseries 22},  262 (2020)}.

\bibitem{Muga2021}
L.~Qi, J.~Chiaverini, H.~Espinós, M.~Palmero, and J.~Muga, ``{\em Fast and
  robust particle shuttling for quantum science and technology}'',
  \href{http://dx.doi.org/10.1209/0295-5075/134/23001}{EPL {\bfseries 134},
  23001 (2021)}.

\bibitem{Torrontegui2011}
E.~Torrontegui, S.~Ib{\'{a}}{\~{n}}ez, X.~Chen, A.~Ruschhaupt,
  D.~Gu{\'{e}}ry-Odelin, and J.~G. Muga, ``{\em {Fast atomic transport without
  vibrational heating}}'',
  \href{http://dx.doi.org/10.1103/PhysRevA.83.013415}{Phys. Rev. A {\bfseries
  83},  013415 (2011)}.

\bibitem{Muga2009}
J.~G. Muga, X.~Chen, A.~Ruschhaupt, and D.~Gu{\'{e}}ry-Odelin, ``{\em
  {Frictionless dynamics of Bose–Einstein condensates under fast trap
  variations}}'', \href{http://dx.doi.org/10.1088/0953-4075/42/24/241001}{J.
  Phys. B {\bfseries 42},  241001 (2009)}.

\bibitem{Schaff2011}
J.-F. Schaff, X.-L. Song, P.~Capuzzi, P.~Vignolo, and G.~Labeyrie, ``{\em
  {Shortcut to adiabaticity for an interacting Bose-Einstein condensate}}'',
  \href{http://dx.doi.org/10.1209/0295-5075/93/23001}{EPL {\bfseries 93},
  23001 (2011)}.

\bibitem{Torrontegui2012}
E.~Torrontegui, X.~Chen, M.~Modugno, S.~Schmidt, A.~Ruschhaupt, and J.~G. Muga,
  ``{\em {Fast transport of Bose-Einstein condensates}}'',
  \href{http://dx.doi.org/10.1088/1367-2630/14/1/013031}{New J. Phys.
  {\bfseries 14},  013031 (2012)}.

\bibitem{Landau1976}
L.~D. Landau and E.~M. Lifshitz, {\em Mechanics}.
\newblock Butterworth-Heinemann, 3~ed., 1976.

\bibitem{Bowler2012}
R.~Bowler, J.~Gaebler, Y.~Lin, T.~R. Tan, D.~Hanneke, J.~D. Jost, J.~P. Home,
  D.~Leibfried, and D.~J. Wineland, ``{\em {Coherent Diabatic Ion Transport and
  Separation in a Multizone Trap Array}}'',
  \href{http://dx.doi.org/10.1103/PhysRevLett.109.080502}{Phys. Rev. Lett.
  {\bfseries 109},  080502 (2012)}.

\bibitem{Couvert2008}
A.~Couvert, T.~Kawalec, G.~Reinaudi, and D.~Gu{\'{e}}ry-Odelin, ``{\em {Optimal
  transport of ultracold atoms in the non-adiabatic regime}}'',
  \href{http://dx.doi.org/10.1209/0295-5075/83/13001}{EPL {\bfseries 83},
  13001 (2008)}.

\bibitem{Guery2014}
D.~Gu\'ery-Odelin and J.~G. Muga, ``{\em Transport in a harmonic trap:
  Shortcuts to adiabaticity and robust protocols}'',
  \href{http://dx.doi.org/10.1103/PhysRevA.90.063425}{Phys. Rev. A {\bfseries
  90},  063425 (2014)}.

\bibitem{Reichle2006}
R.~Reichle, D.~Leibfried, R.~Blakestad, J.~Britton, J.~Jost, E.~Knill,
  C.~Langer, R.~Ozeri, S.~Seidelin, and D.~Wineland, ``{\em {Transport dynamics
  of single ions in segmented microstructured Paul trap arrays}}'',
  \href{http://dx.doi.org/10.1002/prop.200610326}{Fortschr. Phys. {\bfseries
  54},  666--685 (2006)}.

\bibitem{MartinezCercos2020}
D.~Martínez-Cercós, D.~Guéry-Odelin, and J.~G. Muga, ``{\em Robust load
  transport by an overhead crane with respect to cable length uncertainties}'',
  \href{http://dx.doi.org/10.1177/1077546319899204}{J. Vib. Control {\bfseries
  26},  1514-1522 (2020)}.

\bibitem{Lam2021}
M.~R. Lam, N.~Peter, T.~Groh, W.~Alt, C.~Robens, D.~Meschede, A.~Negretti,
  S.~Montangero, T.~Calarco, and A.~Alberti, ``{\em Demonstration of quantum
  brachistochrones between distant states of an atom}'', Phys. Rev. X
  {\bfseries 11},  011035 (2021).

\bibitem{Zhang2016}
Q.~Zhang, J.~G. Muga, D.~Gu{\'{e}}ry-Odelin, and X.~Chen, ``{\em Optimal
  shortcuts for atomic transport in anharmonic traps}'',
  \href{http://dx.doi.org/10.1088/0953-4075/49/12/125503}{J. Phys. B {\bfseries
  49},  125503 (2016)}.

\bibitem{Goldberg1989}
D.~E. Goldberg, {\em Genetic Algorithms in Search, Optimization and Machine
  Learning}.
\newblock Addison-Wesley, Boston, 1989.

\bibitem{Pontryagin}
L.~S. Pontryagin, V.~G. Boltyanskii, R.~V. Gamkrelidze, and E.~F. Mishechenko,
  {\em The Mathematical Theory of Optimal Processes}.
\newblock Interscience Publishers, New York, 1962.

\bibitem{Chen2011}
X.~Chen, E.~Torrontegui, D.~Stefanatos, J.-S. Li, and J.~G. Muga, ``{\em
  {Optimal trajectories for efficient atomic transport without final
  excitation}}'', \href{http://dx.doi.org/10.1103/PhysRevA.84.043415}{Phys.
  Rev. A {\bfseries 84},  043415 (2011)}.

\end{thebibliography}\endgroup
\bibliographystyle{rsta}





\end{document}